\documentclass[a4paper,11pt]{article}
\pdfoutput=1 % to allow compilation in JCAP
\usepackage{jcappub}
\usepackage{amsmath}
\usepackage{amssymb}
\usepackage{bm}
\usepackage{graphicx}
\usepackage{enumitem}
\usepackage{geometry}
\usepackage{xspace}
\usepackage{pdflscape}
\usepackage{enumerate}

\makeatletter
\gdef\@fpheader{}
\g@addto@macro\bfseries{\boldmath}
\makeatother

\newcommand{\ie}{{i.e.~}}

\newcommand{\eg}{e.g.~}

\let\oldsqrt\sqrt
\def\sqrt{\mathpalette\DHLhksqrt}
\def\DHLhksqrt#1#2{%
\setbox0=\hbox{$#1\oldsqrt{#2\,}$}\dimen0=\ht0
\advance\dimen0-0.2\ht0
\setbox2=\hbox{\vrule height\ht0 depth -\dimen0}%
{\box0\lower0.4pt\box2}}

\newcommand{\order}[1]{\mathcal{O}\!\left(#1\right)}

\newcommand{\dd}{\mathrm{d}}
\newcommand{\ee}{e}

\newcommand{\sss}[1]{{\scriptscriptstyle{#1}}}

\newcommand{\uPl}{\mathrm{Pl}}
\newcommand{\uin}{\mathrm{in}}

\newcommand{\umax}{\mathrm{max}}
\newcommand{\uend}{\mathrm{end}}

\newcommand{\ucl}{\mathrm{cl}}

\newcommand{\uc}{\mathrm{c}}

\newcommand{\lo}{\sss{\mathrm{LO}}}
\newcommand{\nlo}{\sss{\mathrm{NLO}}}

\newcommand{\usssPl}{\sss{\uPl}}

\newcommand{\calP}{\mathcal{P}}

\newcommand{\Mp}{M_\usssPl}

\newcommand{\efolds}{$e$-folds}

\newcommand{\beq}{\begin{equation}}
\newcommand{\eeq}{\end{equation}}
\newcommand{\bea}{\begin{equation}\begin{aligned}}
\newcommand{\eea}{\end{aligned}\end{equation}}

\newlength{\wsingfig}
\newlength{\wdblefig}
\newlength{\wquadfig}
\newlength{\wtriplefig}
\newcommand{\Eq}[1]{Eq.~(\ref{#1})}
\newcommand{\Eqs}[1]{Eqs.~(\ref{#1})}
\newcommand{\Fig}[1]{Fig.~{\ref{#1}}}
\newcommand{\Figs}[1]{Figs.~{\ref{#1}}}
\newcommand{\Refa}[1]{Ref.~{\cite{#1}}}
\newcommand{\Refs}[1]{Refs.~{\cite{#1}}}
\newcommand{\Sec}[1]{Sec.~\ref{#1}}
\newcommand{\Secs}[1]{Secs.~\ref{#1}}
\newcommand{\App}[1]{Appendix~\ref{#1}}

\subheader{}

\title{Ultra-slow-roll inflation with quantum diffusion}

\author[a]{Chris Pattison,}
\author[b,a]{Vincent Vennin,}
\author[a]{David Wands,}
\author[a,c]{and Hooshyar Assadullahi}

\affiliation[a]{Institute of Cosmology \& Gravitation, University of Portsmouth, Dennis Sciama Building, Burnaby Road, Portsmouth, PO1 3FX, United Kingdom}
\affiliation[b]{Laboratoire Astroparticule et Cosmologie, CNRS Universit\'e de Paris, 75013 Paris, France}
\affiliation[c]{School of Mathematics and Physics, University of Portsmouth, Lion Gate Building, Lion Terrace, Portsmouth, PO1 3HF, United Kingdom}

\emailAdd{christopher.pattison@port.ac.uk}
\emailAdd{vincent.vennin@port.ac.uk}
\emailAdd{david.wands@port.ac.uk}
\emailAdd{hooshyar.assadullahi@port.ac.uk}

\date{today}

\begin{document}

\sloppy

\abstract{We consider the effect of quantum diffusion on the dynamics of the inflaton during a period of ultra-slow-roll inflation. We extend the stochastic-$\delta\mathcal{N}$ formalism to the ultra-slow-roll regime and show how this system can be solved analytically in both the classical-drift and quantum-diffusion dominated limits. 
By deriving the characteristic function, we are able to construct the full probability distribution function for the primordial density field. In the diffusion-dominated limit, we recover an exponential tail for the probability distribution, as found previously in slow-roll inflation.
To complement these analytical techniques, we present numerical results found both by very large numbers of simulations of the Langevin equations, and through a new, more efficient approach based on iterative Volterra integrals.
We illustrate these techniques with two examples of potentials that exhibit an ultra-slow-roll phase leading to the possible production of primordial black holes.
}


% \arxivnumber{XXXX.XXXXX}

\maketitle

\section{Introduction}
\label{sec:intro}
Cosmological inflation \cite{Starobinsky:1980te, Sato:1980yn, Guth:1980zm, Linde:1981mu, Albrecht:1982wi, Linde:1983gd} is a phase of accelerated expansion that took place at high energy in the primordial universe. It explains its homogeneity and isotropy at large scales. Furthermore, during inflation, quantum vacuum fluctuations are amplified and swept up by the expansion to seed the observed large-scale structure of our universe~\cite{Mukhanov:1981xt, Mukhanov:1982nu, Starobinsky:1982ee, Guth:1982ec, Hawking:1982cz, Bardeen:1983qw}.

Those fluctuations can be described with the curvature perturbation, $\zeta$, and at scales accessible to Cosmic Microwave Background (CMB) experiments~\cite{Ade:2015xua, Ade:2015lrj}, they are constrained to be small ($\zeta\sim 10^{-5}$ until they re-enter the Hubble radius). At smaller scales however, $\zeta$ may become much larger, such that,  upon re-entry to the Hubble radius, it can overcome the pressure gradients and collapse into primordial black holes (PBHs)~\cite{Hawking:1971ei,Carr:1974nx,Carr:1975qj}. This is why the renewed interest in PBHs triggered by the LIGO--Virgo detections~\cite{LIGOScientific:2018mvr}, see \eg \Refs{Clesse:2020ghq, Abbott:2020mjq}, has led to a need to understand inflationary fluctuations produced in non-perturbative regimes. 

Observational constraints on the abundance of PBHs are usually stated in terms of their mass fraction $\beta_M$, defined such that $\beta_M\dd\ln M$  corresponds to the fraction of the mass density of the universe comprised in PBHs of masses between $M$ and $M+\dd\ln M$, at the time PBHs form. Light PBHs ($10^9\mathrm{g}< M < 10^{16} \mathrm{g}$) are mostly constrained from the effects of Hawking evaporation on big bang nucleosynthesis and the extragalactic photon background, and constraints typically range from $\beta_M<10^{-24}$ to $\beta_M<10^{-17}$. Meanwhile, heavier PBHs ($10^{16} \mathrm{g} < M < 10^{50} \mathrm{g}$) are constrained by dynamical or gravitational effects (such as the microlensing of quasars for instance), with $\beta_M<10^{-11}$ to $\beta_M<10^{-5}$ (see \Refa{Carr:2020gox} for a recent review of constraints).

%The standard 
A simple calculation of the abundance of PBHs proceeds as follows. In the perturbative regime, curvature fluctuations produced during inflation have a Gaussian distribution function $P(\zeta)$, where the width of the Gaussian is simply related to the power spectrum of $\zeta$. The mass fraction then follows from the probability that the mean value of $\zeta$ inside a Hubble patch exceeds a certain formation threshold $\zeta_\uc$,  
\bea
\label{eq:def:beta}
\beta_M \sim 2\int_{\zeta_\uc}^\infty P\left(\zeta\right) \dd \zeta\, .
\eea
This formula follows from the Press-Schechter formalism~\cite{1974ApJ...187..425P}, while more refined estimates can be obtained in the excursion-set~\cite{Peacock:1990zz, Bower:1991kf, Bond:1990iw} or peak-theory~\cite{Bardeen:1985tr} approaches. In practice, $\zeta_\uc$ depends on the details of the formation process but is typically of order one~\cite{Zaballa:2006kh, Harada:2013epa},\footnote{The density contrast can also provide an alternative criterion for the formation of PBHs, see \Refa{Young:2014ana}.} and $M$ is some fraction of the mass contained in a Hubble patch at the time of formation~\cite{Choptuik:1992jv, Niemeyer:1997mt, Kuhnel:2015vtw}. If $P(\zeta)$ is Gaussian, $\beta_M$ is thus directly related to the power spectrum of $\zeta$, and the only remaining task is to compute this power spectrum in different inflationary models. 

However, it has recently been pointed out~\cite{Pattison:2017mbe, Biagetti:2018pjj, Ezquiaga:2018gbw} that since PBHs require large fluctuations to form, a perturbative description may not be sufficient. Indeed, primordial curvature perturbations are expected to be well-described by a quasi-Gaussian distribution only when they are small and close enough to the maximum of their probability distribution. Large curvature perturbations on the other hand, which may be far from the peak of the distribution, are affected by the presence of the non-Gaussian tails associated with the unavoidable quantum diffusion of the field(s) driving inflation~\cite{Pattison:2017mbe, Ezquiaga:2019ftu}. In order to describe those tails, one thus requires a non-perturbative approach, such as the stochastic-$\delta N$ formalism~\cite{Enqvist:2008kt, Fujita:2013cna, Vennin:2015hra} that we will employ in this work. Stochastic inflation~\cite{Starobinsky:1982ee, Starobinsky:1986fx} is an effective infrared (low energy, large scale) theory that treats the ultraviolet (high energy, small scale) modes as a source term that provides quantum kicks to the classical motion of the fields on super-Hubble scales. Combined with the $\delta N$ formalism~\cite{Starobinsky:1982ee, Starobinsky:1986fxa, Sasaki:1995aw, Wands:2000dp, Lyth:2004gb}, it provides a framework in which curvature perturbations are related to fluctuations in the integrated expansion and can be characterised from the stochastic dynamics of the fields driving inflation.

The reason why quantum diffusion plays an important role in shaping the mass fraction of PBHs is that, for large cosmological perturbations to be produced, the potential of the fields driving inflation needs to be sufficiently flat, hence the velocity of the fields induced by the potential gradient is small and easily taken over by quantum diffusion. Another consequence of having a small gradient-induced velocity is that it may also be taken over by the velocity inherited from previous stages during inflation where the potential is steeper. This is typically the case in potentials featuring inflection points~\cite{Garcia-Bellido:2017mdw, Germani:2017bcs}.  When this happens, inflation proceeds in the so-called ultra-slow roll (USR) (or ``friction dominated'') regime~\cite{Inoue:2001zt, Kinney:2005vj}, which can be stable under certain conditions~\cite{Pattison:2018bct}, and our goal is to generalise and make use of the stochastic-$\delta N$ formalism in such a phase. 

In practice, we consider the simplest realisation of inflation, where the acceleration of the universe is driven by a single scalar field called the inflaton, $\phi$, whose classical motion is given by the Klein-Gordon equation in a Friedmann-Lemaitre-Robertson-Walker spacetime
\bea \label{eq:kleingordon}
\ddot{\phi} + 3H\dot{\phi} + V'(\phi) = 0 \, ,
\eea 
where $H$ is the Hubble rate, $V(\phi)$ is the potential energy of the field, and a prime denotes a derivative with respect to the field $\phi$ while a dot denotes a derivative with respect to cosmic time $t$. In the usual slow-roll approximation, the field acceleration can be neglected and \Eq{eq:kleingordon} reduces to $3H\dot{\phi} \simeq -V'(\phi)$. In this limit, the stochastic formalism has been shown to have excellent agreement with usual quantum field theoretic (QFT) techniques in regimes where the two approaches can be compared~\cite{Starobinsky:1994bd, Tsamis:2005hd, Finelli:2008zg, Garbrecht:2013coa, Vennin:2015hra, Onemli:2015pma, Burgess:2015ajz, Vennin:2016wnk, Hardwick:2017fjo, Tokuda:2017fdh}. However the stochastic approach can go beyond perturbative QFT using the full nonlinear equations of general relativity to describe the non-perturbative evolution of the coarse-grained field. This can be used to reconstruct the primordial density perturbation on super-Hubble scales using the stochastic-$\delta N$ approach mentioned above. In particular this approach has recently been used to reconstruct the full probability density function (PDF) for the primordial density field~\cite{Pattison:2017mbe,Ezquiaga:2019ftu}, finding large deviations from Gaussian statistics in the nonlinear tail of the distribution. 
More precisely, these tails were found to be exponential rather than Gaussian, which cannot be properly described by the usual, perturbative parametrisations of non-Gaussian statistics (such as those based on computing the few first moments of the distribution and the non-linearity parameters $f_{\mathrm{NL}}$, $g_{\mathrm{NL}}$, etc.), which can only account for polynomial modulations of Gaussian tails~\cite{Byrnes:2012yx, Passaglia:2018ixg, Atal:2018neu}.

When the inflaton crosses a flat region of the potential, such that $|V'|\ll 3H|\dot\phi|$, the slow-roll approximation breaks down. In the limit where the potential gradient can be neglected, \ie the USR limit, one has $\ddot{\phi} + 3H\dot{\phi} \simeq 0$, which leads to $\dot{\phi} = \dot{\phi}_{\mathrm{in}}\ee^{-3N}$ where $N=\int H\, \dd t$ is the number of \efolds. 
If in addition we consider the de-Sitter limit where $H$ is constant, we obtain
\bea
\label{eq:usr:dotphisol}
\dot{\phi}=\dot{\phi}_\uin+3H(\phi_\uin-\phi)
\,, 
\eea
so the field's classical velocity is simply a linear function of the field value. 
Notice that the USR evolution is thus dependent on the initial values of both the inflaton field and its velocity, unlike in slow roll where the slow-roll attractor trajectory is independent of the initial velocity. 

Let us note that stochastic effects in USR have been already studied by means of the stochastic-$\delta N$ formalism in \Refs{Firouzjahi:2018vet, Ballesteros:2020sre} (see also \Refs{Cruces:2018cvq, Prokopec:2019srf} for an alternative approach), which investigated leading corrections to the first statistical moments of the curvature perturbation. Our work generalises these results by addressing the full distribution function of curvature perturbations, including in regimes where quantum diffusion dominates, and by drawing conclusions for the production of PBHs from a (stochastic) USR period of inflation. Let us also mention that a numerical analysis has recently been performed in \Refa{1836208}, in the context of a modified Higgs potential, where the non-Markovian effect of quantum diffusion on the amplitude of the noise itself was also taken into account. Our results complement this work, by providing analytical insight from simple toy models.  

This paper is organised as follows. In \Sec{sec:StochasticInflation} we review the stochastic inflation formalism and apply it to the USR setup. We explain how the full PDF of the integrated expansion, $N$, given the initial field value and its velocity, can be computed from first-passage time techniques. In practice, it obeys a second-order partial differential equation that we can solve analytically in two limits, that we study in \Secs{sec:classicallimit} and~\ref{sec:stochasticlimit}. In  \Sec{sec:classicallimit}, we construct a formal solution for the characteristic function of the PDF in the regime where quantum diffusion plays a subdominant role compared with the classical drift (\ie the inherited velocity). We recover the results of \Refs{Firouzjahi:2018vet} in this case. Then, in \Sec{sec:stochasticlimit}, we consider the opposite limit where quantum diffusion is the main driver of the inflaton's dynamics. We recover the results of \Refa{Pattison:2017mbe} at leading order where the classical drift is neglected, and we derive the corrections induced by the finite classical drift at next-to-leading order. In both regimes, we compare our analytic results against numerical realisations of the Langevin equations driving the stochastic evolution. Since such a numerical method is computationally expensive (given that it relies on solving a very large number of realisations, in order to properly sample the tails of the PDF), in \Sec{sec:volterra} we outline a new method, that makes use of Volterra integral equations, and which, in some regimes of interest, is far more efficient than the Langevin approach. Finally, in \Sec{sec:applications} we apply our findings to two prototypical toy-models featuring a USR phase, and we present our conclusions in \Sec{sec:conclusions}. The paper ends with two appendices where various technical aspects of the calculation are deferred.
\section{The stochastic-$\delta N$ formalism for ultra-slow roll inflation}
\label{sec:StochasticInflation}
As mentioned in \Sec{sec:intro}, hereafter we consider the framework of single-field inflation, but our results can be straightforwardly extended to multiple-field setups~\cite{Assadullahi:2016gkk}.
\subsection{Stochastic inflation beyond slow roll}
Away from the slow-roll attractor, the full phase space dynamics needs to be described, since the homogeneous background field $\phi$ and its conjugate momentum $\pi\equiv\dot\phi/H$ are two independent dynamical variables. Following the techniques developed in \Refa{Grain:2017dqa}, in the Hamiltonian picture, \Eq{eq:kleingordon} can be written as two coupled first-order differential equations
\begin{align} 
\label{eq:conjmomentum} \frac{\dd {\phi}}{\dd N} &= {\pi} \, ,\\
\label{eq:KG:efolds} \frac{\dd{\pi}}{\dd N} &= - \left(3-\epsilon_{1}\right){\pi} - \frac{V'({\phi})}{H^2}  \, ,
\end{align}
where 
$\epsilon_{1}\equiv - \dot{H}/H^2$ is the first slow-roll parameter. The Friedmann equation 
\bea 
\label{eq:friedmann}
H^2 = \dfrac{V+\frac{1}{2}\dot{\phi}^2}{3\Mp^2}\, ,
\eea 
where $\Mp$ is the reduced Planck mass, provides a relationship between the Hubble rate, $H$, and the phase-space variables $\phi$ and $\pi$.\footnote{In terms of $\phi$ and $\pi$, one has $H^2=V(\phi)(3\Mp^2-\pi^2/2)^{-1}$ and $\epsilon_1=\pi^2/(2\Mp^2)$.\label{footnote:H_eps1:phi_pi}}

The stochastic formalism is an effective theory for the long-wavelength parts of the quantum fields during inflation, so we split the phase-space fields into a long-wavelength part and an inhomogeneous component that accounts for linear fluctuations on small-scales,
\begin{align}
\hat{\phi} &= \hat{\bar{\phi}} + \hat{\phi}_{\mathrm{s}} \label{eq:decomp:phi}\, , \\
\hat{\pi} &= \hat{\bar{\pi}} + \hat{\pi}_{\mathrm{s}} \label{eq:decomp:pi}\, ,
\end{align}
where $\hat{\phi}$ and $\hat{\pi}$ are now quantum operators, which we make explicit with the hats.
The short-wavelength parts of the fields, $\hat{\phi}_{\mathrm{s}}$ and $\hat{\pi}_{\mathrm{s}}$, can be written as 
\begin{align}
    \hat{\phi}_{\mathrm{s}} &= \int_{\mathbb{R}^3}\frac{\dd \bm{k}}{\left(2\pi\right)^{\frac{3}{2}}}W\left( \frac{k}{\sigma aH}\right) \left[ \ee^{-i\bm{k}\cdot\bm{x}}\phi_{k}(N)\hat{a}_{\bm{k}} + \ee^{i\bm{k}\cdot\bm{x}}\phi_{k}^{*}(N)\hat{a}^{\dagger}_{\bm{k}}  \right] \, ,\\
        \hat{\pi}_{\mathrm{s}} &= \int_{\mathbb{R}^3}\frac{\dd \bm{k}}{\left(2\pi\right)^{\frac{3}{2}}}W\left( \frac{k}{\sigma aH}\right) \left[ \ee^{-i\bm{k}\cdot\bm{x}}\pi_{k}(N)\hat{a}_{\bm{k}} + \ee^{i\bm{k}\cdot\bm{x}}\pi_{k}^{*}(N)\hat{a}^{\dagger}_{\bm{k}}  \right] \, .
\end{align}
In these expressions $\hat{a}^{\dagger}_{\bm{k}}$ and $\hat{a}_{\bm{k}}$ are creation and annihilation operators, and $W$ denotes a window function that selects out modes with  $k> k_\sigma$ where $k_\sigma = \sigma a H$ is the comoving coarse-graining scale and $\sigma\ll 1$ is the coarse-graining parameter. 
The coarse-grained fields $\bar{\phi}$ and $\bar{\pi}$ thus contain all wavelengths that are much larger than the Hubble radius, $k< k_\sigma$. 
These long-wavelength components are the local background values of the fields, which are continuously perturbed by the small wavelength modes as they cross the coarse-graining radius. 

By inserting the decompositions \eqref{eq:decomp:phi} and \eqref{eq:decomp:pi} into the classical equations of motion \eqref{eq:conjmomentum} and \eqref{eq:KG:efolds}, to linear order in the short-wavelength parts of the fields, the equations for the long-wavelength parts are found to be
\begin{align}
\frac{\partial \hat{\bar{\phi}}}{\partial N} &= \hat{\bar{\pi}} + \hat{\xi}_{\phi}(N) \label{eq:conjmomentum:langevin:quantum} \, ,\\
\frac{\partial \hat{\bar{\pi}}}{\partial N} &= -\left(3-\epsilon_{1}\right)\hat{\bar{\pi}} - \frac{V'(\hat{\bar{\phi}})}{H^2} +\hat{\xi}_{\pi}(N) \label{eq:KG:efolds:langevin:quantum} \, ,
\end{align}
where the source functions $\hat{\xi}_{\phi}$ and $\hat{\xi}_{\pi}$ are given by
\begin{align}
    \hat{\xi}_{\phi} &= -\int_{\mathbb{R}^3}\frac{\dd \bm{k}}{\left(2\pi\right)^{\frac{3}{2}}}\frac{\dd W}{\dd N}\left( \frac{k}{\sigma aH}\right) \left[ \ee^{-i\bm{k}\cdot\bm{x}}\phi_{k}(N)\hat{a}_{\bm{k}} + \ee^{i\bm{k}\cdot\bm{x}}\phi_{k}^{*}(N)\hat{a}^{\dagger}_{\bm{k}}  \right]\, , \\
    \hat{\xi}_{\pi} &= -\int_{\mathbb{R}^3}\frac{\dd \bm{k}}{\left(2\pi\right)^{\frac{3}{2}}}\frac{\dd W}{\dd N}\left( \frac{k}{\sigma aH}\right) \left[ \ee^{-i\bm{k}\cdot\bm{x}}\pi_{k}(N)\hat{a}_{\bm{k}} + \ee^{i\bm{k}\cdot\bm{x}}\pi_{k}^{*}(N)\hat{a}^{\dagger}_{\bm{k}}  \right] \, .
\end{align}
If the window function $W$ is taken as a Heaviside function, the two-point correlation functions of the sources are given by
\bea 
\langle 0| \hat{\xi}_{\phi}(N_1)\hat{\xi}_{\phi}(N_2)|0\rangle & 
\kern-0.1em = \kern-0.1em
 \frac{1}{6\pi^2}\frac{\dd k_\sigma^3(N)}{\dd N}\bigg|_{N_1}\left\vert\phi_{k_\sigma}(N_1) \right\vert^2 \delta\left(N_1 - N_2\right) \kern-0.1em ,\\
\langle 0| \hat{\xi}_{\pi}(N_1) \hat{\xi}_{\pi}(N_2)|0\rangle &
\kern-0.1em = \kern-0.1em
\frac{1}{6\pi^2}\frac{\dd k_\sigma^3(N)}{\dd N}\bigg|_{N_1}\left\vert \pi_{k_\sigma}(N_1) \right\vert^2 \delta\left(N_1 - N_2\right) \kern-0.1em, \\
\langle 0| \hat{\xi}_{\phi}(N_1)\hat{\xi}_{\pi}(N_2)|0\rangle &
\kern-0.1em = \kern-0.1em \langle 0| \hat{\xi}_{\pi}(N_1) \hat{\xi}_{\phi}(N_2)|0\rangle^* 
\kern-0.1em = \kern-0.1em
 \frac{1}{6\pi^2}\frac{\dd k_\sigma^3(N)}{\dd N}\bigg|_{N_1}\phi_{k_\sigma}(N_1) \pi^*_{k_\sigma}(N_1)  \delta\left(N_1 \kern-0.1em - \kern-0.1em N_2\right)  \kern-0.1em .
 \label{eq:noise:correlators}
\eea 
In particular, for a massless field in a de-Sitter background, in the long-wavelength limit, $\sigma\ll 1$, one obtains~\cite{Grain:2017dqa}
\bea 
 \langle 0| \hat{\xi}_{\phi}(N_1)\hat{\xi}_{\phi}(N_2)|0\rangle  
&\simeq
 \left( \frac{H}{2\pi} \right)^2
 \delta\left(N_1 - N_2\right) \,  ,\\
 \langle 0| \hat{\xi}_{\pi}(N_1) \hat{\xi}_{\pi}(N_2)|0\rangle 
 & \simeq
 0 \, ,\\
 \langle 0| \hat{\xi}_{\phi}(N_1)\hat{\xi}_{\pi}(N_2)|0\rangle 
 & = \kern-0.1em \langle 0| \hat{\xi}_{\pi}(N_1) \hat{\xi}_{\phi}(N_2)|0\rangle^* 
 \simeq
0 \, .
 \label{eq:dS:correlators}
\eea 

The next step is to realise that on super-Hubble scales, once the decaying mode becomes negligible, commutators can be dropped and the quantum field dynamics can be cast in terms of a stochastic system~\cite{Lesgourgues:1996jc, Martin:2015qta, Grain:2017dqa, Vennin:2020kng}. In this limit, the source functions can be interpreted as random Gaussian noises rather than quantum operators, which are correlated according to \Eqs{eq:noise:correlators}. The dynamical equations~\eqref{eq:conjmomentum:langevin:quantum} and~\eqref{eq:KG:efolds:langevin:quantum} can thus be seen as stochastic Langevin equations for the random field variables $\bar{\phi}$ and $\bar{\pi}$, 
\begin{align}
\frac{\partial {\bar{\phi}}}{\partial N} &= {\bar{\pi}} + {\xi}_{\phi}(N) \label{eq:conjmomentum:langevin} \, ,\\
\frac{\partial {\bar{\pi}}}{\partial N} &= -\left(3-\epsilon_{1}\right){\bar{\pi}} - \frac{V'({\bar{\phi}})}{H^2(\bar{\phi},\bar{\pi})} +{\xi}_{\pi}(N) \label{eq:KG:efolds:langevin} \, ,
\end{align} 
where we have removed the hats to stress that we now work with classical stochastic quantities rather than quantum operators. Note that since the time coordinate is not perturbed in \Eqs{eq:conjmomentum:langevin}-\eqref{eq:KG:efolds:langevin}, the Langevin equations are implicitly derived in the uniform-$N$ gauge, so the field fluctuations must be computed in that gauge when evaluating \Eq{eq:noise:correlators}~\cite{Pattison:2019hef}. The stochastic formalism also relies on the separate universe approach, which allows us to use the homogeneous equations of motion to describe the inhomogeneous field at leading order in a gradient expansion, and which has been shown to be valid even beyond the slow-roll attractor in \Refa{Pattison:2019hef}.

\subsection{Stochastic ultra-slow-roll inflation}
\label{sec:Stochastic:USR}
\begin{figure}
    \centering
    \includegraphics[width=0.6\textwidth]{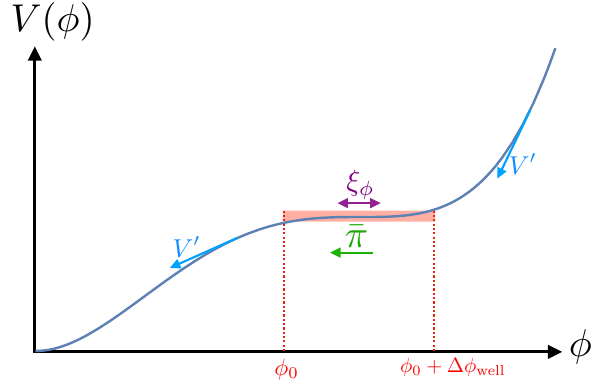}
    \caption{Sketch of the setup considered in this work. The inflationary potential features a flat region, between $\phi_0$ and $\phi_0+\Delta\phi_{\mathrm{well}}$, where the potential gradient can be neglected and inflation proceeds in the ultra slow roll regime, \ie $\phi$ decreases because of the velocity $\bar{\pi}$ it has acquired in previous stages. Outside this range, the field is driven by the potential gradient, $V'$, in the slow-roll regime. Stochastic effects also make the inflaton fluctuate inside the USR well, with a noise amplitude given by $H/(2\pi)$. The dynamics inside the USR well follow the Langevin equations~\eqref{eq:eom:phi:stochastic}-\eqref{eq:eom:v:stochastic}. 
    \label{fig:sketch}}
\end{figure}

In the absence of a potential gradient ($V'\simeq 0$) and in the quasi de-Sitter limit ($\epsilon_1\simeq 0$), plugging \Eqs{eq:dS:correlators} into \Eqs{eq:conjmomentum:langevin} and~\eqref{eq:KG:efolds:langevin} lead to the Langevin equations 
\begin{align}
\label{eq:eom:phi:stochastic} 
\frac{\dd \bar{\phi}}{\dd N} &= \bar{\pi}+ \frac{H}{2\pi}\xi(N) \, , \\
\frac{\dd \bar{\pi}}{\dd N} &= -3\bar{\pi} 
\label{eq:eom:v:stochastic} \, ,
\end{align}
where $\xi(N)$ is Gaussian white noise with unit variance, such that $\langle\xi(N)\rangle = 0$ and $\langle \xi(N)\xi(N')\rangle=\delta(N-N')$. In practice, in a given inflationary potential, ultra slow roll takes place only across a finite field range, which we denote $[\phi_0,\phi_0+ \Delta\phi_\mathrm{well}]$ and which we refer to as the ``USR well''. Outside this range, we assume that the potential gradient is sufficiently large to drive a phase of slow-roll inflation. A sketch of this setup is displayed in \Fig{fig:sketch}.
It is convenient to rewrite \Eqs{eq:eom:phi:stochastic} and~\eqref{eq:eom:v:stochastic} in terms of rescaled, dimensionless variables. A relevant value of the field velocity is the one for which the classical drift is enough to cross the USR well. Making use of \Eq{eq:usr:dotphisol}, it is given by the value $\dot{\phi}_\uin$ such that $\phi$ reaches $\phi_0$ as $\dot{\phi}$ reaches zero, hence 
\bea
\label{eq:critical:velocity}
\bar{\pi}_{\mathrm{cri}}=-3\Delta\phi_{\mathrm{well}}\, .
\eea 
Since the typical field range of the problem is $\Delta\phi_{\mathrm{well}}$, we thus introduce 

\bea
 \label{eq:xy:variabletransforms}
x = \frac{\bar{\phi} - \phi_0}{\Delta \phi_\mathrm{well}} \, ,  \quad
y = \frac{\bar{\pi}}{\bar{\pi}_\mathrm{crit}} \, .
\eea 
Upon crossing the USR well, $x$ varies from $1$ to  $0$, and $y$, which is positive, decays from its initial value $y_\uin$. If $y_\uin>1$, the field would cross the USR well by means of the classical velocity only, while if $y_\uin<1$, in the absence of quantum diffusion, it would come to a stop at $x=1-y_\uin$. We therefore expect stochastic effects to play a crucial role when $y_\uin<1$.  In terms of $x$ and $y$, the Langevin equations~\eqref{eq:eom:phi:stochastic}-\eqref{eq:eom:v:stochastic} become
\bea
 \label{eq:langevin:xy}
\frac{\dd x}{\dd N} &= -3 y + \frac{\sqrt{2}}{\mu} \xi(N)\, , \\
\frac{\dd y}{\dd N} &= -3y \, ,
\eea 
where the dimensionless parameter
\bea \label{eq:def:mu}
\mu \equiv 2\sqrt{2}\pi \frac{\Delta \phi_\mathrm{well}}{H}
\eea 
has been introduced, following the notations of \Refs{Pattison:2017mbe,Ezquiaga:2019ftu} and in order to make easy the comparison with those works. The two physical parameters of the problem are therefore $y_\uin$, which depends on the slope of the potential at the entry point of the USR well, and $\mu$, which describes the width of the quantum well relative to the Hubble rate, and which controls the amplitude of stochastic effects. Note that, in the absence of quantum diffusion, according to \Eq{eq:langevin:xy}, $x-y$ is a constant. Let us also stress that since $\epsilon_1=\bar{\pi}^2/2$, see footnote~\ref{footnote:H_eps1:phi_pi}, in order for inflation not to come to an end before entering the USR well, one must impose $\epsilon_1<1$, which yields an upper bound on $y_\uin$, namely
\bea
 \label{eq:ymax:generic}
y_\mathrm{in} < \frac{\sqrt{2}\Mp}{3\Delta\phi_\mathrm{well}} \, .
\eea
In particular, if one considers the case of a super-Planckian well, $\Delta\phi_\mathrm{well} \gg \Mp$, then the above condition imposes that the initial velocity is close to zero, and one recovers the limit studied in \Refa{Pattison:2017mbe} where the effect of the inherited velocity $\pi$ was neglected.

Finally, the Langevin equations~\eqref{eq:langevin:xy} give rise to a Fokker--Planck equation that drives the probability density in phase space at time $N$, $P(x,y; N)$, which evolves according to~\cite{risken1989fpe}
\bea \label{eq:FP:USR}
\frac{\partial P(x,y;N)}{\partial N} &= \left[ 3 + 3y\left(\frac{\partial}{\partial x} + \frac{\partial}{\partial y}\right) + \frac{1}{\mu^2}\frac{\partial^2}{\partial x^2}\right] P (x,y;N)
% \\
&\equiv \mathcal{L}_{\mathrm{FP}} \cdot P(x,y;N) \, .
\eea 
This equation defines the Fokker--Planck operator $\mathcal{L}_{\mathrm{FP}}$, which is a differential operator in phase space.
\subsection{First-passage time problem}
\label{sec:}
In the $\delta N$ formalism~\cite{Starobinsky:1982ee, Starobinsky:1986fxa, Sasaki:1995aw, Wands:2000dp, Lyth:2004gb}, the curvature perturbation on large scales is given by the fluctuation in the integrated expansion $N$, between an initial flat hypersurface and a final hypersurface of uniform energy density. Our goal is therefore to compute the number of \efolds~spent in the toy model depicted in \Fig{fig:sketch}, which is a random variable that we denote $\mathcal{N}$, and the statistics of $\delta\mathcal{N}=\mathcal{N}-\langle \mathcal{N} \rangle$ will give access to the statistics of the curvature perturbation $\zeta$ (hereafter, ``$\langle\cdot\rangle$'' denotes ensemble average over realisations of the Langevin equations). Note that if stochastic effects are subdominant in the slow-roll parts of the potential (between which the USR well is sandwiched, see \Fig{fig:sketch}), then the contribution of the slow-roll parts to $\mathcal{N}$ is only to add a constant, hence $\delta\mathcal{N}$ is given by the fluctuation in the number of \efolds~spent in the USR well only, which is why we focus on the USR regime below.

The PDF for the random variable $\mathcal{N}$, starting from a given $x$ and $y$ in phase space, is denoted $P(\mathcal{N};x,y)$ [not to be confused with $P(x,y;N)$ introduced in \Eq{eq:FP:USR}]. In \Refs{Vennin:2015hra, Pattison:2017mbe, Ezquiaga:2019ftu}, it is shown to obey the adjoint Fokker-Planck equation,
\bea
\label{eq:adjoint:FP:equation:PDF}
\frac{\partial P(\mathcal{N};x,y)}{\partial \mathcal{N}} = \mathcal{L}_\mathrm{FP}^\dagger \cdot P(\mathcal{N};x,y)
=\left[ \frac{1}{\mu^2}\frac{\partial^2}{\partial x^2} - 3y\left(\frac{\partial}{\partial x} + \frac{\partial}{\partial y}\right) \right]P(\mathcal{N};x,y) .
\eea
In this expression, $\mathcal{L}_\mathrm{FP}^\dagger$ is the adjoint Fokker-Planck operator, related to the Fokker-Planck operator via $\int \dd x\dd y F_1(x,y) \mathcal{L}_{\mathrm{FP}}\cdot F_2(x,y) = \int \dd x \dd y F_2(x,y) \mathcal{L}_{\mathrm{FP}}^\dagger\cdot F_1(x,y) $.  The adjoint Fokker-Planck equation is a partial differential equation in 3 dimensions ($x,y$ and $\mathcal{N}$), which can be cast in terms of a partial differential equation in 2 dimensions by Fourier transforming the $\mathcal{N}$ coordinate. This can be done by introducing the characteristic function~\cite{Pattison:2017mbe}
\bea 
\label{eq:def:charfunction}
\chi_{\mathcal{N}}(t; x,y) &= \left\langle e^{it\mathcal{N}(x,y)}\right\rangle
=\int_{-\infty}^{\infty}e^{it\mathcal{N}}
P(\mathcal{N};x,y)
 \dd \mathcal{N},
\eea 
where $t$ is a dummy parameter. From the second equality, one can see that the characteristic function is nothing but the Fourier transform of the PDF, hence the PDF can be obtained by inverse Fourier transforming the characteristic function,
\bea
\label{eq:PDF:chi}
P\left(\mathcal{N}; x, y\right) = \frac{1}{2\pi} \int^{\infty}_{-\infty} \ee^{-it\mathcal{N}} \chi_{\mathcal{N}}\left(t; x, y \right)\dd t\, .
\eea
Let us also note that by Taylor expanding the exponential function in \Eq{eq:def:charfunction}, the moments of the PDF are given by
\bea
\label{eq:meanefolds:charfunction}
\left\langle\mathcal{N}^n\right\rangle = i^{-n}\frac{\partial^n}{\partial t^n}\left. \chi_{\mathcal{N}}(t; x,y) \right\vert_{t=0}\, .
\eea
By plugging \Eq{eq:PDF:chi} into \Eq{eq:adjoint:FP:equation:PDF}, the characteristic function is found to obey
\bea
\label{eq:diff:chi}
\left(\mathcal{L}_{\mathrm{FP}}^\dagger  +it\right)\chi_\mathcal{N}\left(t;x,y\right)= 0\, ,
\eea
which is indeed a partial differential equation in two dimensions ($x$ and $y$).\footnote{Since \Eq{eq:diff:chi} is linear in $x$, one can further reduce it down to a partial differential equation in one dimension, \ie an ordinary differential equation, by Fourier transforming $x$. This however makes the boundary conditions~\eqref{eq:char:initialconditions} difficult to enforce and does not bring much analytical insight.} It needs to be solved in the presence of an absorbing boundary at $x=0$ (if quantum diffusion is negligible in the lower slow-roll region, once the field has exited the well, it follows the gradient of the potential and cannot return to the USR well) and a reflective boundary at $x=1$ (if quantum diffusion is negligible in the upper slow-roll region, the potential gradient prevents the field from visiting that region once in the USR well), \ie
\bea
 \label{eq:char:initialconditions}
\chi_\mathcal{N}(t; 0, y) = 1 \, , \quad \frac{\partial \chi_\mathcal{N}}{\partial x}(t; 1, y) = 0 \, .
\eea 
In the following, we solve this equation in the regime where the initial velocity $y_\uin$ is large and quantum diffusion plays a negligible role, see \Sec{sec:classicallimit}, and in the opposite limit where the initial velocity is small and the field is mostly driven by quantum diffusion, see \Sec{sec:stochasticlimit}.  

\section{Drift-dominated regime}
\label{sec:classicallimit}
We first consider the regime where quantum diffusion provides a small contribution to the duration of (the majority of) the realisations of the Langevin equations. As explained in \Sec{sec:Stochastic:USR}, for the field to cross the USR well without the help of quantum diffusion, one must have $y_\uin>1$. A necessary condition is therefore that $y_\uin\gg 1$. As we will see below, one must also impose $\mu y_\uin \gg 1$, which is a stricter requirement in the case where $\mu<1$.
\subsection{Leading order}
At leading order, the term proportional to $\partial^2/\partial x^2$ can be neglected in \Eq{eq:adjoint:FP:equation:PDF}, since it corresponds to quantum diffusion in the $x$ direction, and \Eq{eq:diff:chi} reduces to
\bea
 \label{eq:char:classicallimit}
\left[ - 3y\left(\frac{\partial}{\partial x} + \frac{\partial}{\partial y}\right) + it \right] \left. \chi_\mathcal{N}\right\vert _{\lo} (t;x,y)= 0 \, .
\eea 
This partial differential equation being first order, it can be solved with the method of characteristics, see \App{app:classicalLO:charfunction} where we provide a detailed derivation of the solution. Together with the boundary conditions~\eqref{eq:char:initialconditions}, one obtains\footnote{Note that the solution~\eqref{eq:char:classicalLO:solution} does not respect the reflective boundary condition at $x=1$. This is because, since \Eq{eq:char:classicallimit} is first order, only one boundary condition can be imposed. Indeed, at leading order in the drift-dominated limit, the system follows the classical trajectory in phase space and never bounces against the reflective boundary, which explains why the reflective boundary is irrelevant.}
\bea 
\label{eq:char:classicalLO:solution}
\left. \chi_\mathcal{N}\right\vert _\lo(t;x,y) = \left( 1 - \frac{x}{y}\right)^{-\frac{it}{3}} \, .
\eea 
Making use of \Eq{eq:meanefolds:charfunction}, the mean number of \efolds~at that order is given by
\bea
 \label{eq:meanefolds:classical:LO}
\left.\langle \mathcal{N}\rangle \right\vert_\lo(x,y) 
&= -\frac{1}{3}\ln\left( 1 - \frac{x}{y}\right) \, ,
\eea 
which is nothing but the classical, deterministic result. Furthermore, by Fourier transforming \Eq{eq:char:classicalLO:solution} according to \Eq{eq:PDF:chi}, one obtains for the PDF
\bea 
\label{eq:pdf:cl:lo}
P(\mathcal{N};x,y) = \delta\left[ \mathcal{N} - \left.\langle \mathcal{N}\rangle \right\vert_\lo(x,y) \right] \, ,
\eea 
where $\delta$ denotes the Dirac distribution, and which confirms that at that order, all realisations follow the same, classical trajectory.
\subsection{Next-to-leading order}
At next-to-leading order, the term proportional to $\partial^2/\partial x^2$ in \Eq{eq:diff:chi} can be evaluated with the leading-order solution~\eqref{eq:char:classicalLO:solution}, which leads to
\bea
\label{eq:Drift_dominated:nlo:rec}
\left[ - 3y\left(\frac{\partial}{\partial x} + \frac{\partial}{\partial y}\right) + it \right]\left.\chi_\mathcal{N}\right\vert_\nlo(t;x,y) = -  \frac{1}{\mu^2}\frac{\partial^2}{\partial x^2} \left.\chi_{\mathcal{N}}\right\vert_\lo (t;x,y)\, .
\eea  
This partial differential equation is still of first order and can again be solved with the methods of characteristics, and in \App{app:classicalLO:charfunction} one obtains
\bea \label{eq:characteristic:NLO}
\chi_\mathcal{N}\bigg|_\mathrm{NLO} &= \left( 1 - \frac{x}{y}\right)^{-\frac{it}{3}}\left[ 1 - \frac{it}{9}\left(1+\frac{it}{3}\right)\frac{\ln\left( 1-\frac{x}{y}\right)}{\mu^2(y-x)^2} \right] 
\, .
\eea 
It is interesting to note that contrary to the slow-roll case where the characteristic function is the one of a Gaussian distribution at next-to-leading order~\cite{Pattison:2017mbe} (at leading order, it is simply a Dirac distribution following the classical path) and starts deviating from Gaussian statistics only at next-to-next-to leading order, in USR the PDF differs from a Gaussian distribution already at next-to-leading order. 

The mean number of \efolds~can again be calculated using \Eq{eq:meanefolds:charfunction}, which leads to\footnote{The NLO correction found in \Refa{Firouzjahi:2018vet}
[see equation $(4.15)$ in that reference] 
is given by $({\kappa^2}/{6}) \left< \mathcal{N} \right> \big|_\lo$, where,

in the notation of our paper,
$\kappa = -({\sqrt{2}}/{\mu})(y-x)^{-1}$.
We therefore recover the same result, despite using different techniques.
}
\bea \label{eq:meaefolds:classical:NLO}
\langle \mathcal{N}\rangle(x,y)\Big|_{\mathrm{NLO}} 
&= \left< \mathcal{N} \right> \big|_\lo \left[ 1 + \frac{1}{3\mu^2\left(y-x\right)^2} \right] \, .
\eea 
The relative correction to the leading-order result is controlled by $(\mu y)^{-2}$ (recalling that $y\gg 1>x$), so as announced above, the present expansion also requires that $\mu y\gg 1$.

One can also calculate the power spectrum~\cite{Fujita:2013cna, Ando:2020fjm}
\bea 
\label{eq:Pzetageneral}
P_{\zeta} = \frac{\dd \langle \delta \mathcal{N}^2 \rangle}{\dd \langle \mathcal{N}\rangle} 
\, ,
\eea 
where $\langle \delta \mathcal{N}^2 \rangle= \langle\mathcal{N}^2\rangle - \langle\mathcal{N}\rangle^2$, which relies on computing the second moment of $\mathcal{N}$. At leading order, combining \Eqs{eq:meanefolds:charfunction} and~\eqref{eq:char:classicalLO:solution}, one simply has $\langle \mathcal{N}^2\rangle\vert_\lo = \langle \mathcal{N} \rangle\vert_\lo^2$, so one has to go to next-to-leading order where \Eqs{eq:meanefolds:charfunction} and~\eqref{eq:characteristic:NLO} give rise to
\bea 
\langle\mathcal{N}^2\rangle_\nlo
&= 
\left\langle \mathcal{N} \right\rangle \big|_\nlo^2
+ 
\frac{2\left\langle \mathcal{N} \right\rangle \big|_\lo}{9\mu^2(y-x)^2}
\, .
\eea 
The leading contribution to $ \langle \delta \mathcal{N}^2 \rangle$ is thus given by
\bea 
\label{eq:deltaN2:cl:nlo}
 \langle \delta \mathcal{N}^2 \rangle &= \frac{2}{9\mu^2(y-x)^2}\left< \mathcal{N} \right> \big|_\lo
 \, . \\
\eea 
Note that, as mentioned above, $y-x=$ is constant along the classical trajectory. Therefore, in the classical limit, \Eq{eq:Pzetageneral} yields\footnote{At higher order, we expect the power spectrum to be affected by stochastic modifications of the trajectories along which the derivatives in \Eq{eq:Pzetageneral} are evaluated~\cite{Ando:2020fjm}.}
\bea 
P_{\zeta}(x,y) \simeq \frac{2}{9\mu^2\left(y-x\right)^2} \, ,
\eea 
which coincides with the usual perturbative calculation of the power spectrum in ultra-slow roll \cite{Firouzjahi:2018vet}. One can also note that the condition to be in the classical regime, $\mu y\gg 1$, implies that the power spectrum remains small.

\subsection{Series expansion for large velocity}

The procedure outlined above can be extended to a generic expansion in inverse powers of $\mu^2 y^2$,
\bea
\label{eq:chi:large:muy:expansion}
\chi_{\mathcal{N}} (t;x,y) = 
\sum_{n=0}^\infty \left( \mu^2 y^2 \right)^{-n} C_n \left( t;\frac{x}{y} \right) \ .
\eea
The boundary condition $\chi_\mathcal{N}(t;x,y)= 1$ at $x=0$ is zeroth-order in $\mu y$ and thus requires $C_0(t;0)=1$ and $C_n(t;0)=0$ for all $n\geq1$.

The lowest-order solution $C_0(t;x/y)$ obeys the classical limit of the equation for the characteristic function, \Eq{eq:char:classicallimit} and the solution is given by \Eq{eq:char:classicalLO:solution}. 
The higher-order functions $C_n(t;u)$ for $n\geq1$ obey the recurrence relation
\bea
\label{eq:classicalrecurrence}
(1-u)C_{n+1}^\prime - \left[ \frac{it}{3}+2(n+1) \right] C_{n+1} = \frac13 C_{n}^{\prime\prime} \,,
\eea
where a prime denotes derivatives with respect to the argument $u$, and which was obtained by plugging \Eq{eq:chi:large:muy:expansion} into \Eq{eq:diff:chi}.
Thus we have an integrating-factor type first-order differential equation for each successive function, $C_{n+1}(t;u)$, determined by the second derivative of the preceding function, $C_n(t;u)$.
The solutions are of the form\footnote{Note that we require $y\geq x$ and hence $u=x/y\leq 1$ for the classical trajectory to cross the potential well and reach $x=0$.}
\bea
\label{eq:Cn:cnm}
C_n(t;u) = \sum_{m=0}^n c_{n,m}(t) \left( 1-u \right)^{-\frac{it}{3}-2n} \left[ \ln \left( 1-u \right) \right]^m \, ,
\eea
where the coefficients $c_{n,m}(t)$ are polynomial functions of order up to $2n$, given by the recurrence relations (for $1\leq m \leq n+1$)
\bea
\label{eq:cn:iterative}
c_{n+1,m}(t)  
&= - \frac{1}{3m} \left[ \left( \frac{it}{3} + 2n \right) \left( \frac{it}{3} +2n  + 1 \right) c_{n,m-1}(t) 
\right. \\ & \qquad \qquad \left.
-2m \left( \frac{it}{3} + 2n +\frac{1}{2}\right) c_{n,m}(t) 
+ m(m+1) c_{n,m+1}(t) 
\right] \,.
\eea
Here, 
$c_{0,0}=1$ and $c_{n,0}=0$ for all $n\geq1$, and we set $c_{n,m}=0$ for all $m>n$. This allows one to compute the $c_{n,m}$ coefficients iteratively, and the first coefficients are given by
\bea
\label{eq:cnm:first}
& c_{0,0}(t) = 1
\,, \\
& c_{1,0}(t) = 0 \,, \quad
 c_{1,1}(t) = - \frac{it}{9} \left( 1+ \frac{it}{3} \right)
\,, \\
& c_{2,0}(t) = 0 \,, \quad
c_{2,1}(t) = \frac{c_{1,1}(t)}{3} \left( \frac{2it}{3} +5 \right) \,, \quad
c_{2,2}(t) = - \frac{c_{1,1}(t)}{6} \left( \frac{it}{3} + 2\right) \left( \frac{it}{3} +3 \right)
\,.
\eea

It is straightforward to implement \Eq{eq:cn:iterative} numerically and to compute the characteristic function, hence the PDF, to arbitrary order in that expansion. For instance, the higher-order corrections to the mean number of \efolds~are found to be
\bea
\langle \mathcal{N}\rangle   
& =
\langle\mathcal{N}\rangle_\lo  \left[1+\frac{1}{3 \mu ^2 (x-y)^2}+\frac{9 \langle \mathcal{N}\rangle_\lo+5}{9 \mu ^4 (x-y)^4}+\frac{60 \langle \mathcal{N}\rangle_\lo^2+77 \langle \mathcal{N}\rangle_\lo+17}{9 \mu ^6 (x-y)^6}
\right. \\ & \left. \qquad \qquad  \qquad  
+\frac{5670 \langle \mathcal{N}\rangle_\lo^3+12042 \langle \mathcal{N}\rangle_\lo^2+6396 \langle \mathcal{N}\rangle_\lo+817}{81 \mu ^8 (x-y)^8}+\cdots\right] .
\eea 
One can check that this expression again coincides with the one given up to order $\mu^{-6}(x-y)^{-6}$ in \Refa{Firouzjahi:2018vet}. Here we obtain it as the result of a systematic expansion~\eqref{eq:chi:large:muy:expansion} that can be performed to arbitrary order without further calculations:
\bea
\langle \mathcal{N}\rangle   
& = \sum_{n=0}^\infty \sum_{m=0}^n 
\left[ -ic_{n,m}'(0) - \frac{c_{n,m}(0)}{3} \ln \left(1-\frac{x}{y}\right) \right] \frac{\left[ \ln \left( 1-\frac{x}{y} \right) \right]^m}{\mu^{2n}(x-y)^{2n}}
\,.
\eea

The non-Gaussian nature of the PDF close to its maximum can be characterised by the local non-linearity parameter
\bea
\label{eq:fnl}
f_{\mathrm{NL}}=\frac{5}{36 \calP_\zeta^2}\frac{\dd^2 \left\langle \delta\mathcal{N}^3\right\rangle}{\dd \left\langle \mathcal{N}\right\rangle ^2}\, .
\eea
In this expression, the third moment can be obtained by combining \Eqs{eq:Cn:cnm} and~\eqref{eq:cnm:first}, keeping terms up to $n=2$ in \Eq{eq:chi:large:muy:expansion}. Then \Eq{eq:meanefolds:charfunction} for the third moment gives
\bea
\left\langle \delta \mathcal{N}^3 \right\rangle = \left\langle \left( \mathcal{N} - \left\langle \mathcal{N} \right\rangle \right)^3\right\rangle = \frac{4 \left[\log \left(1-\frac{x}{y}\right)-1\right] \log \left(1-\frac{x}{y}\right)}{81 \mu ^4 (x-y)^4}\, .
\eea
This yields
\bea
f_{\mathrm{NL}} = \frac{5}{2}+\order{\mu^{-2}y^{-2}}\, ,
\eea 
which is of order one, while it is suppressed by the slow-roll parameters in the slow-roll regime~\cite{Maldacena:2002vr}. This expression also matches the standard USR result of \Refa{Namjoo:2012aa}.

\section{Diffusion-dominated regime} 
\label{sec:stochasticlimit}
In this section, we consider the opposite limit where the initial velocity, inherited from the phase preceding the USR epoch, is small, and the field is mostly driven by the stochastic noise. In the same way that the drift-dominated regime was studied through a $1/y$-expansion, the diffusion-dominated regime can be approached via a $y$-expansion. Such a systematic expansion is performed in \App{app:separable:charfunction}. Here, we only derive the result at leading and next-to-leading orders, before discussing implications for primordial black hole production.

Since the classical drift decays exponentially with time, $y = y_\mathrm{in}\ee^{-3N}$, it always becomes negligible at late time. It is thus important to stress that the diffusion-dominated regime becomes effective at late time, so the results derived in this section are relevant for the upper tail of the PDF of $\mathcal{N}$, even if the initial velocity is substantial. Since PBHs precisely form in that tail, this limit is therefore important to study the abundance of these objects.
 \begin{figure} 
    \centering
    \includegraphics[width=0.49\textwidth]{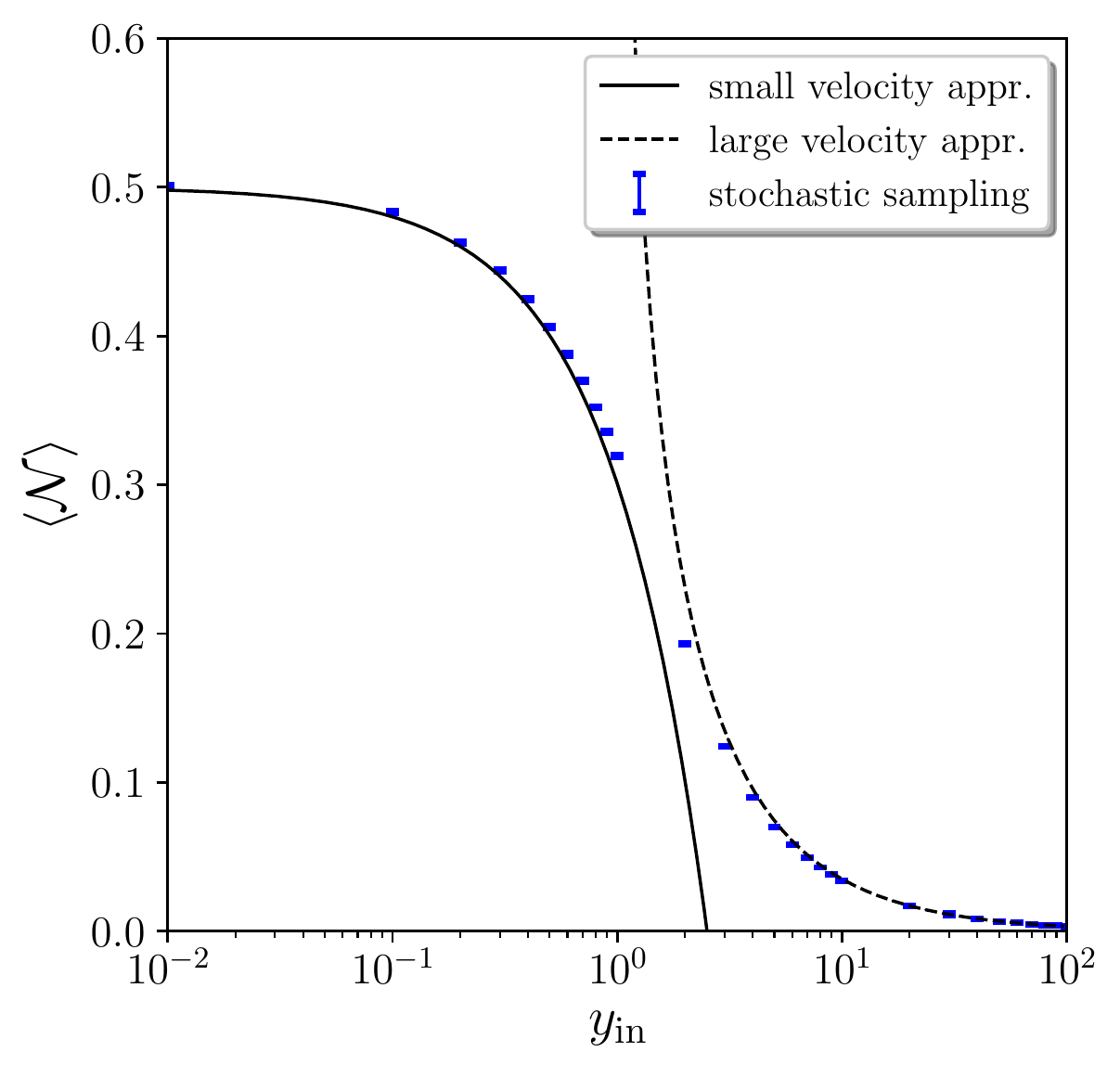}
    \includegraphics[width=0.49\textwidth]{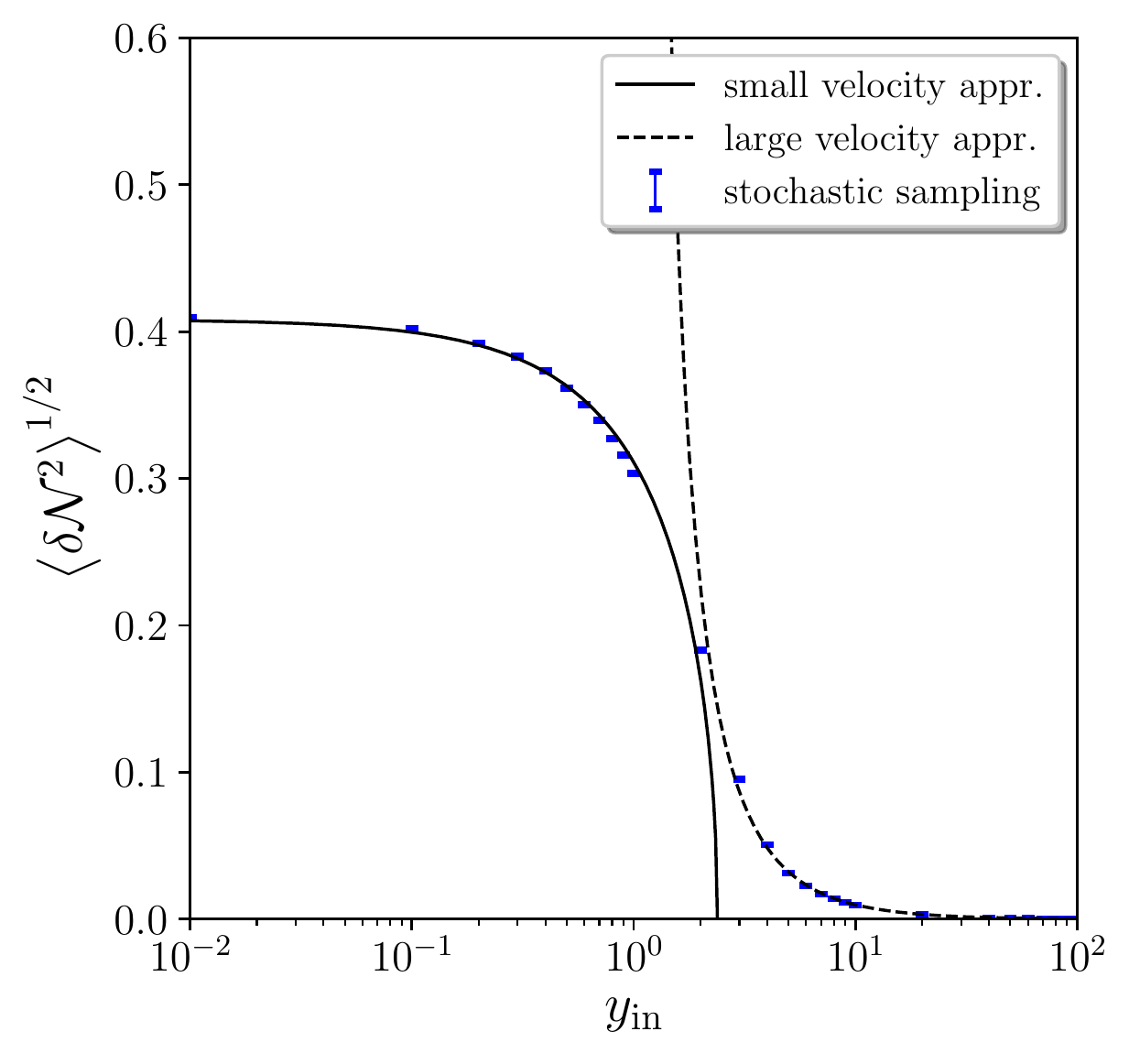}
    \caption{Mean number of \efolds~(left panel) across the USR well, and its standard deviation (right panel), as a function of the rescaled initial velocity $y_\uin$, for $\mu=1$. The blue bars are reconstructed from a large number (between $10^6$ and $10^8$, depending on the value of $y_\uin$) of realisations of the Langevin equations \eqref{eq:langevin:xy}. The size of the bars correspond to a $2\sigma$-estimate for the statistical error, which is obtained using the jackknife resampling method (see footnote~\ref{footnote:jackknife}).
    The dashed black curve corresponds to the drift-dominated, leading-order result derived in \Sec{sec:classicallimit} while the solid black lines display the diffusion-dominated, next-to-leading-order result obtained in \Sec{sec:stochasticlimit}.
    \label{fig:meanefolds:smally}}
\end{figure}
\subsection{Leading order} 
\label{sec:latetimelimit}
At leading order, one can simply set $y=0$ in the characteristic equation~\eqref{eq:diff:chi}, given explicitly in \Eq{eq:pde:chi:xy}, and one has 
\bea \label{eq:char:latetime}
\frac{1}{\mu^2}\frac{\partial^2}{\partial x^2}\chi_{\mathcal{N}}(t; x, 0) = -it\chi_{\mathcal{N}}(t;x,0) \, .
\eea 
This describes free diffusion on a flat potential where the inflaton has no classical velocity. In this zero-velocity limit, we recover the \textit{slow-roll} Langevin equations for a flat potential, which was previously studied as a limit of slow-roll inflation in \Refa{Pattison:2017mbe}. 
Classically this limit is not well-defined, since the inflaton remains at rest and inflation never ends if there is no classical velocity, but when one accounts for quantum diffusion the problem is ``regularised", giving a finite duration of inflation for a finite field range.

The solution to \Eq{eq:char:latetime} that obeys the boundary conditions~\eqref{eq:char:initialconditions} is given by~\cite{Pattison:2017mbe,Ezquiaga:2019ftu}
\bea \label{eq:chi:usr:latetime}
\chi_{\mathcal{N}}(t;x,0) = 
 \frac{\cos\left[ \omega_0\left(1-x\right)\right]}{\cos\omega_0}
\, ,
\eea 
where we used the notation introduced in \App{app:separable:charfunction},
\bea 
\label{eq:omega0}
\omega_0^2 = i t \mu^2 
\, . 
\eea 
This allows one to compute the mean number of \efolds~from \eqref{eq:meanefolds:charfunction} for instance, and one finds 
\bea \label{eq:usr:efolds:analytic}
\left. \left< \mathcal{N}(x)\right> \right|_{y=0}= \mu^2x\left(1-\frac{x}{2}\right) \,  ,
\eea 
or the second moment, given by
\bea
\label{eq:usr:efolds:2ndmoment:analytic}
\left. \left< \delta\mathcal{N}^2(x)\right> \right|_{y=0}=\frac{2+x(x-2)}{3}\mu^2 \left. \left< \mathcal{N}(x)\right> \right|_{y=0}\, .
\eea 
By comparing this expression with \Eq{eq:deltaN2:cl:nlo}, one can see that the way it scales with $\mu$ is different than in the drift-dominated regime: in the drift-dominated regime, the typical size of the fluctuation decreases with $\mu$, while it increases with $\mu$ is the diffusion-dominated regime. When $\mu\ll 1$ the transition between the two regimes occurs in the range $1\ll y\ll \mu$ where neither approximation applies, and where the size of the typical fluctuation thus smoothly increases with $y$. If $\mu\gg 1$, the transition occurs at $y\sim 1$ and is therefore more abrupt. 

The full PDF of the number of \efolds~is given by the inverse Fourier transform of the characteristic function, see \eqref{eq:PDF:chi}, which leads to 
\bea \label{eq:pdf:usr:analyitic}
P\left(\mathcal{N}; x, 0 \right) = & \frac{2 \pi}{\mu^2} \sum_{n=0}^\infty \left( n + \frac{1}{2} \right) 
 \sin\left[\left(n+\frac{1}{2}\right) \pi x \right] \exp\left[ -\frac{\pi^2}{\mu^2} \left(n+\frac{1}{2}\right)^2 \mathcal{N}\right] \\
= & -\frac{\pi}{2 \mu^2} \vartheta_{2}' \left( \frac{\pi}{2}x, \ee^{-\frac{\pi^2}{\mu^2} \mathcal{N}} \right) \, ,
\eea 
where $\vartheta_{2}$ is the second elliptic theta function~\cite{Abramovitz:1970aa:theta} 
\bea 
\vartheta_2\left(z,q\right) &= 2 \sum_{n=0}^\infty  q^{\left(n+\frac{1}{2}\right)^2}\cos\left[\left(2n+1\right)z\right]\, ,
\eea 
and $\vartheta_2^\prime$ is its derivative with respect to the first argument. One can check that this distribution is properly normalised and that its first moment is given by \Eq{eq:usr:efolds:analytic}. When $\mathcal{N}$ is large, the mode $n=0$ dominates in the sum of \Eq{eq:pdf:usr:analyitic}, which thus features an exponential, heavy tail. Those exponential tails have important consequences for PBHs, which precisely form from them~\cite{Pattison:2017mbe, Ezquiaga:2019ftu}. 

As explained above, since the small-velocity limit is a late-time limit, we expect the decay rate of the leading exponential, $\ee^{-\Lambda_0 \mathcal{N}}$ with $\Lambda_0=\pi^2/(4\mu^2)$, to remain the same even if the initial velocity does not vanish, since the realisations of the Langevin process that contribute to the tail of the PDF escape the USR well at a time when the velocity has decayed away.
\subsection{Next-to-leading order}
\label{sec:stochlimit:expansion}
At next-to-leading order, the term proportional to $y$ in the characteristic equation~\eqref{eq:diff:chi} [given explicitly in \Eq{eq:pde:chi:xy}] can be evaluated with the leading-order solution. Formally, by expanding the characteristic function as 
\bea 
\label{eq:linear:chi}
\chi_\mathcal{N}(t;x,y) \approx \chi_\mathcal{N}(t;x,0) + yf(t;x) \, , 
\eea 
where $\chi_\mathcal{N}(t;x,0)$ is given by \Eq{eq:chi:usr:latetime}, the characteristic equation reads
\bea 
\left[ \frac{1}{\mu^2}\frac{\partial^2}{\partial x^2} + \left( it - 3 \right) \right]f(t;x) &= 3\frac{\partial}{\partial x}\chi_\mathcal{N}(t;x,0)
= 3\omega_0 \frac{\sin\left[\omega_0\left(1-x\right)\right]}{\cos\omega_0}
\, , 
\eea 
where only the terms linear in $y$ have been kept. This equation needs to be solved with the boundary conditions~\eqref{eq:char:initialconditions}, which require
\bea \label{eq:boundaryconditions:smallylimit} 
& f(t;0) = 0 \, , & \frac{\partial f}{\partial x}(t;1) = 0 \, .
\eea 
One obtains
\bea 
\label{eq:solution:ftx}
f(t;x) &= 
A_1\cos\left[\omega_1\left(1-x\right)\right] + B_1\sin\left[\omega_1\left(1-x\right)\right] 
- \frac{\omega_0\sin\left[\omega_0\left(1-x\right)\right]}{\cos\omega_0}
\, , 
\eea 
where \bea 
\omega_1^2 = \left( it - 3 \right) \mu^2 
\, ,
\eea 
and $A_1$ and $B_1$ are determined by the boundary conditions \eqref{eq:boundaryconditions:smallylimit},
\bea
A_1 &= \frac{\omega_0}{\omega_1} \left( \frac{\omega_1\sin\omega_0 - \omega_0\sin\omega_1}{\cos\omega_0\cos\omega_1} \right) 
 \,,
\hspace{1cm} 
B_1 =  \frac{\omega_0^2}{\omega_1\cos\omega_0} \,.
\eea

As before, we can compute the mean number of \efolds~by using \Eq{eq:meanefolds:charfunction}, and one obtains 
\bea \label{eq:meanefolds:smally}
\left< \mathcal{N} \right>(x,y) &\simeq
\mu^2x\left(1-\frac{x}{2}\right) 
 + \mu^2 y \left\{ x-1 + \frac{\cosh\left[ \sqrt{3}\mu\left(1-x\right) \right]}{\cosh\left( \sqrt{3}\mu \right)} - \frac{\sinh\left( \sqrt{3}\mu x \right) }{\sqrt{3}\mu\cosh\left( \sqrt{3}\mu \right)} \right\} \, ,
\eea 
which reduces to \Eq{eq:usr:efolds:analytic} when $y=0$. 
This expression~\eqref{eq:meanefolds:smally} is displayed as a function of the initial velocity $y_\mathrm{in}$ in the left panel of \Fig{fig:meanefolds:smally}, and compared with the mean number of \efolds~computed over a large number of numerical simulations of the Langevin equations \eqref{eq:eom:phi:stochastic}-\eqref{eq:eom:v:stochastic} for $\mu=1$ and displayed with the blue bars.
The number of numerical realisations that are generated varies between $10^6$ and $10^8$, depending on the value of the initial velocity. The size of the bars correspond to an estimate of the $2\sigma$ statistical error (due to using a finite number of realisations), as obtained from the jackknife resampling method.\footnote{For a sample of $n$ trajectories, the central limit theorem states that the statistical error on this sample scales as $1/\sqrt{n}$, so the standard deviation can be written as $\sigma=\lambda/\sqrt{n}$ for large $n$.
The parameter $\lambda$ can be estimated by dividing the set of realisations into $n_\mathrm{sub}$ subsamples, each of size $n/n_\mathrm{sub}$.
In each subsample, one can compute the mean number of \efolds~$\left<\mathcal{N}\right>_{i}$ where $i=1\cdots n_\mathrm{sub}$, and then compute the standard deviation $\sigma_{n/n_\mathrm{sub}}$ across the set of values of $\left<\mathcal{N}\right>_{i}$. 
This allows one to evaluate $\lambda=\sqrt{n/n_\mathrm{sub}}\sigma_{n/n_\mathrm{sub}}$, and hence $\sigma_n=\sigma_{n_\mathrm{sub}}/\sqrt{n_\mathrm{sub}}$.
In practice, we take $n_\mathrm{sub}=100$.\label{footnote:jackknife}}
One can check that when $y\ll 1$, \Eq{eq:meanefolds:smally} provides a good fit to the numerical result, while when $y\gg 1$, the small-velocity formula, \Eq{eq:meanefolds:classical:LO}, gives a correct description of the numerical result. 

This is also confirmed in the right panel of \Fig{fig:meanefolds:smally}, where the standard deviation of the number of \efolds, $\sqrt{\langle \delta \mathcal{N}^2 \rangle}$, is shown. When $y\ll 1$, the diffusion-dominated result, obtained by plugging \Eq{eq:linear:chi} into \Eq{eq:meanefolds:charfunction} (we do not reproduce the formula here since it is slightly cumbersome and not particularly illuminating) and displayed with the solid black line, provides a good fit to the numerical result, while when $y\gg 1$, the classical result~\eqref{eq:deltaN2:cl:nlo} gives a good approximation.

The above considerations can be generalised into a systematic expansion in $y$ that we set up in \App{app:separable:charfunction}, and which allows one to compute higher corrections up to any desired order.
\subsection{Implications for primordial black hole abundance}
When going from the characteristic function to the PDF via the inverse Fourier transform of \Eq{eq:PDF:chi}, it is convenient to first identify the poles of the characteristic function, given as $\Lambda_n^{(m)} = it$, so that we can expand $\chi_\mathcal{N}$ as~\cite{Ezquiaga:2019ftu}
\bea 
\label{eq:pole:expansion}
\chi_\mathcal{N}(t;x,y) = \sum_{m,n} \frac{a_n^{(m)}(x,y)}{\Lambda_n^{(m)}-it} + g(t;x,y) \, , 
\eea 
where $g(t,x)$ is a regular function, and $a_n^{(m)}(x,y)$ is the residual associated with $\Lambda_n^{(m)}$ and can be found via 
\bea 
\label{eq:residue:chi}
a_n^{(m)}(x,y) = -i\left[\frac{\partial}{\partial t}\chi_\mathcal{N}^{-1}\left(t=-i\Lambda_n^{(m)}; x, y\right)\right]^{-1} \, . 
\eea 
Here, the poles are labeled by two integer numbers $n$ and $m$, for future convenience. By inverse Fourier transforming \Eq{eq:pole:expansion}, one obtains
\bea \label{eq:PDF:pole:expansion}
P\left(\mathcal{N};x,y \right) = \sum_{m,n} a_n^{(m)}(x,y)\ee^{-\Lambda_n^{(m)}\mathcal{N}} \, ,
\eea 
so the location of the poles determine the exponential decay rates and the residues set the (non-necessarily positive) amplitude associated to each decaying exponential. 

At next-to-leading order in the diffusion-dominated limit, the characteristic function is given by \Eq{eq:linear:chi}, where both $\chi_\mathcal{N}(t;x,0)$ and $f(t;x)$ have $\cos\omega_0$ in their denominator, so there is a first series of simple poles when 
$\omega_0=[n+(1/2)]\pi$, where $n$  is an integer number. Making use of \Eq{eq:omega0}, these poles are associated to the decay rates
\bea 
\Lambda_n^{(0)} = \frac{\pi^2}{\mu^2}\left(n+\frac{1}{2}\right)^2  \, . 
\eea 
The corresponding residuals can be obtained by plugging \Eq{eq:linear:chi} into \Eq{eq:residue:chi}, and one obtains
\bea 
a_n^{(0)} 
&=
\frac{(2n+1)\pi}{\mu^2} \left( \sin\left[\left(n+\frac12\right)\pi x\right]
 - y \left(n+\frac12\right)\pi \cos \left[\left(n+\frac12\right)\pi x\right] \right. \\
 & \hspace{1cm} \left. + y \frac{\left(n+\frac12\right)\pi}{\omega_{1,n}^{(0)}\cos\omega_{1,n}^{(0)}}
 \left\lbrace \omega_{1,n}^{(0)} \cos \left[ \omega_{1,n}^{(0)}\left(1-x\right)\right] - (-1)^n \left(n+\frac12\right)\pi \sin\left( \omega_{1,n}^{(0)} x \right) \right\rbrace \right)
\eea 
where 
\bea
\omega_{1,n}^{(0)} \equiv \sqrt{\left(n+\frac12\right)^2\pi^2-3\mu^2} \,.
\eea

The function $f(t;x)$, which multiplies $y$ in the characteristic function \eqref{eq:linear:chi}, also features the term 
$\cos\omega_1$ in its denominator, which has simple poles at $it=\Lambda_n^{(1)}$ where
\bea 
\Lambda_n^{(1)} = 3+ \frac{\pi^2}{\mu^2}\left(n+\frac{1}{2}\right)^2 
\, , 
\eea 
with associated residual at 
\bea 
a_n^{(1)} &= \frac{2(-1)^n y}{\mu^2}\frac{\sin\left[ \left(n+\frac{1}{2}\right)\pi x\right]}{\cos\left[ \sqrt{3\mu^2 + \pi^2\left(n+\frac{1}{2}\right)^2} \right]} \left\{ -3\mu^2 - \pi^2\left(n+\frac{1}{2}\right)^2 \right. \\ 
& \hspace{1cm} \left. +\pi(-1)^n \left(n+\frac{1}{2}\right)\sqrt{3\mu^2 + \pi^2\left( n+\frac{1}{2} \right)^2}\sin\left[ \sqrt{3\mu^2 + \pi^2\left( n+\frac{1}{2} \right)^2} \right] \right\} \, . 
\eea 
The PDF is then obtained from \Eq{eq:PDF:pole:expansion} and reads 
\bea \label{eq:PDF:smallylimit}
P(\mathcal{N}; x, y) &= \sum_{m=0,1}\sum_{n=0}^{\infty}a_n^{(m)}(x,y)\ee^{-\Lambda_n^{(m)}\mathcal{N}} &= \sum_{n=0}^{\infty}\left[ a_n^{(0)} + a_n^{(1)}\ee^{-3\mathcal{N}} \right] \ee^{-\Lambda_n^{(0)}\mathcal{N}} \, ,
\eea
where all quantities appearing in this expression are given above. 

\begin{figure} 
    \centering
    \includegraphics[width=0.49\textwidth]{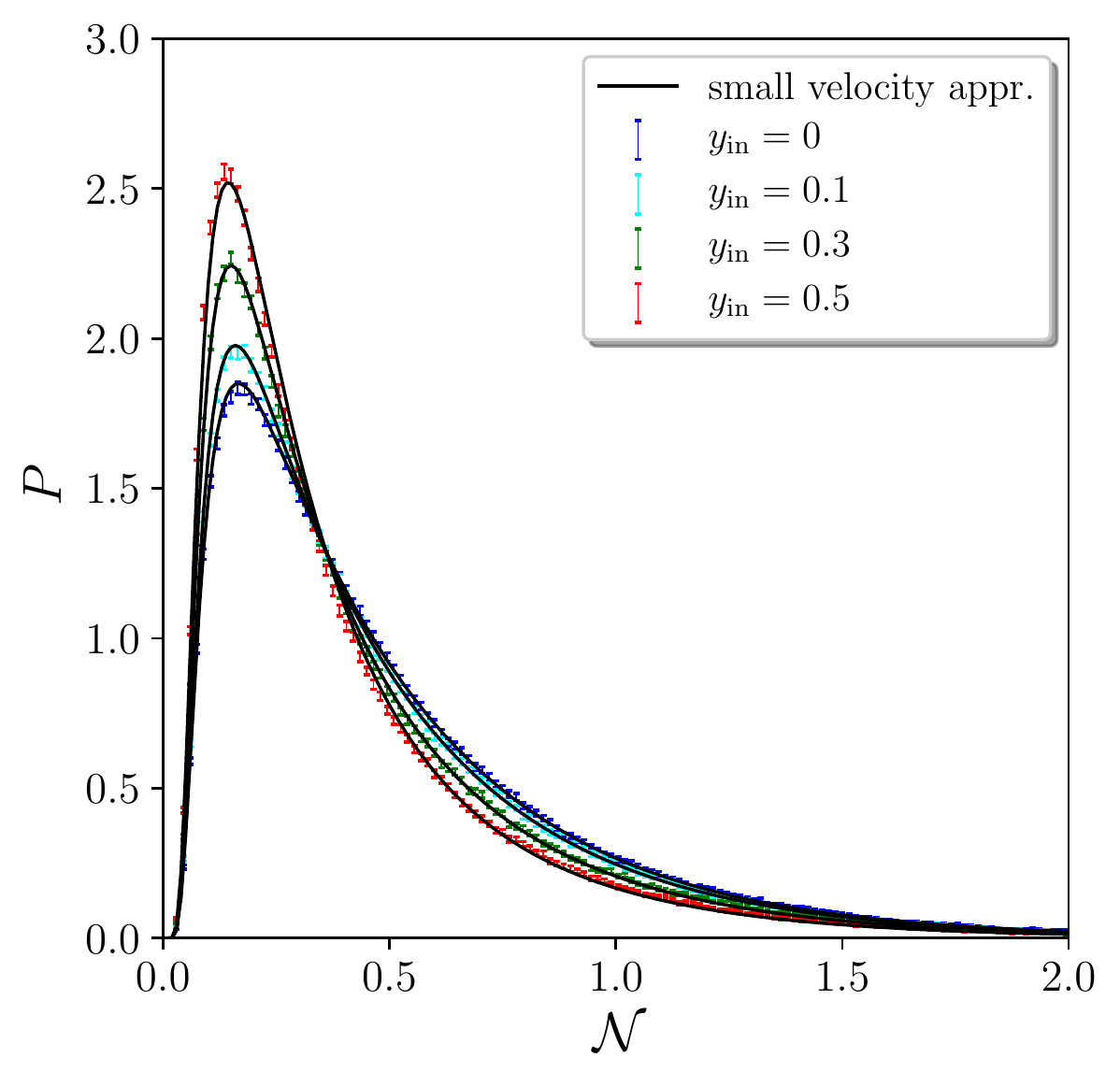}
    \includegraphics[width=0.49\textwidth]{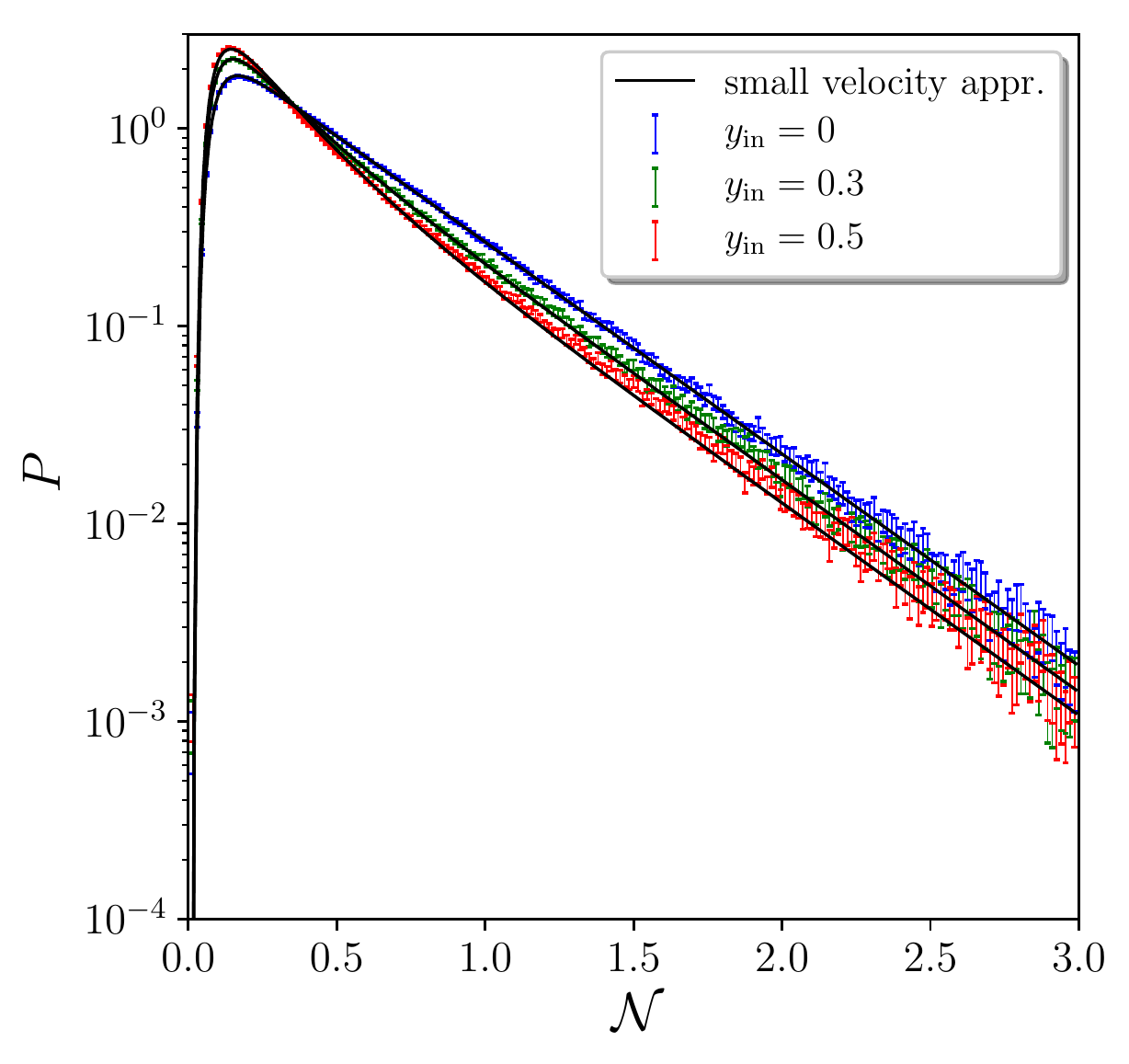}
    \caption{The probability distribution $P$ of the number of \efolds~$\mathcal{N}$ realised through the USR well, for $\mu=1$ and with several values of the initial velocity $y_\mathrm{in}$. In the left panel we use a linear scale of the vertical axis, and in the right panel we use a logarithmic scale (the value $y_\mathrm{in}=0.1$ is not shown in the right panel for display convenience). The black curves correspond to the small-$y$ expression \eqref{eq:PDF:smallylimit} (where in practice, the sum is truncated at $n=1000$), while the error bars are reconstructed from a large number of realisations of the Langevin equations \eqref{eq:langevin:xy}, with Gaussian kernel density of width $\dd N=0.005$. The size of the bars is the $2\sigma$-estimate for the statistical error obtained from the jackknife resampling procedure, see footnote~\ref{footnote:jackknife}. \label{fig:pdf:smally}}
\end{figure}

Note that the presence of the classical velocity does not change the location of the first set of poles, $it=\Lambda_n^{(0)}$, that are already present at leading order in this calculation, but rather adds a second set of poles at $it=\Lambda_n^{(1)}=\Lambda_n^{(0)}+3$. Since the asymptotic behaviour on the tail is given by the lowest pole, \ie $\Lambda_n^{(0)}$, it does not depend on whether or not the classical drift is included, as announced above.\footnote{This is also a consequence of the more generic result shown in \Refa{Ezquiaga:2019ftu} that the poles, hence the decay rates, do not depend on the field-space coordinates (hence on the value of $x$ and $y$ here).} However, the overall amplitude of the leading exponential term, given by $a_0^{(0)}$, does depend on $y$.

Let us also stress that by extending those considerations to the systematic $y$-expansion set up in \App{app:separable:charfunction}, one can show that a new set of poles arises at each order, and given by \Eq{eq:pole:expansion:nm}, \ie $\Lambda_n^{(m)}=\Lambda_n^{(0)}+3m$. Each set of poles is therefore more and more suppressed on the tail, but the sets are well separated only when $\mu\gg 1$. For instance, if $\mu < \sqrt{2/3}\pi$, one has $\Lambda_0^{(1)}<\Lambda_1^{(0)}$, so the second most important term in \Eq{eq:PDF:pole:expansion} comes from the second set of poles, not the first one. 

The formula~\eqref{eq:PDF:smallylimit} is compared to numerical simulations of the Langevin equation in \Fig{fig:pdf:smally}, where we can see that it provides a very good fit to the PDF, even for substantial values of $y_\mathrm{in}$.
On the tail of the distributions, the error bars become larger, and this is because realisations that experience a large number of \efolds~are rare, and hence they provide sparser statistics. They are however sufficient to clearly see, especially in the right panel, that the tails are indeed exponential, with a decay rate that does not depend on $y_\mathrm{in}$.

\begin{figure} 
    \centering
    \includegraphics[width=0.49\textwidth]{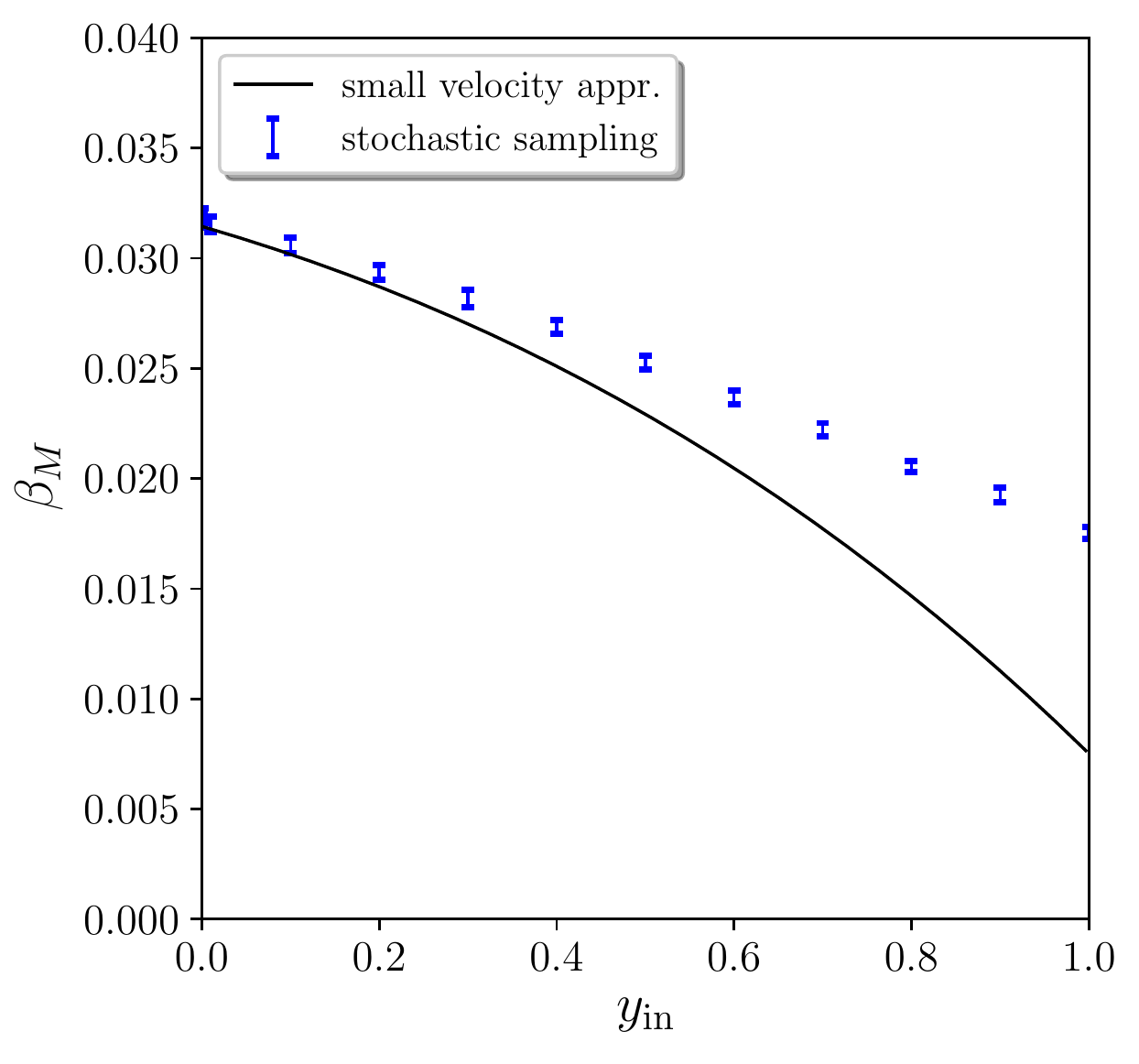}
    \includegraphics[width=0.49\textwidth]{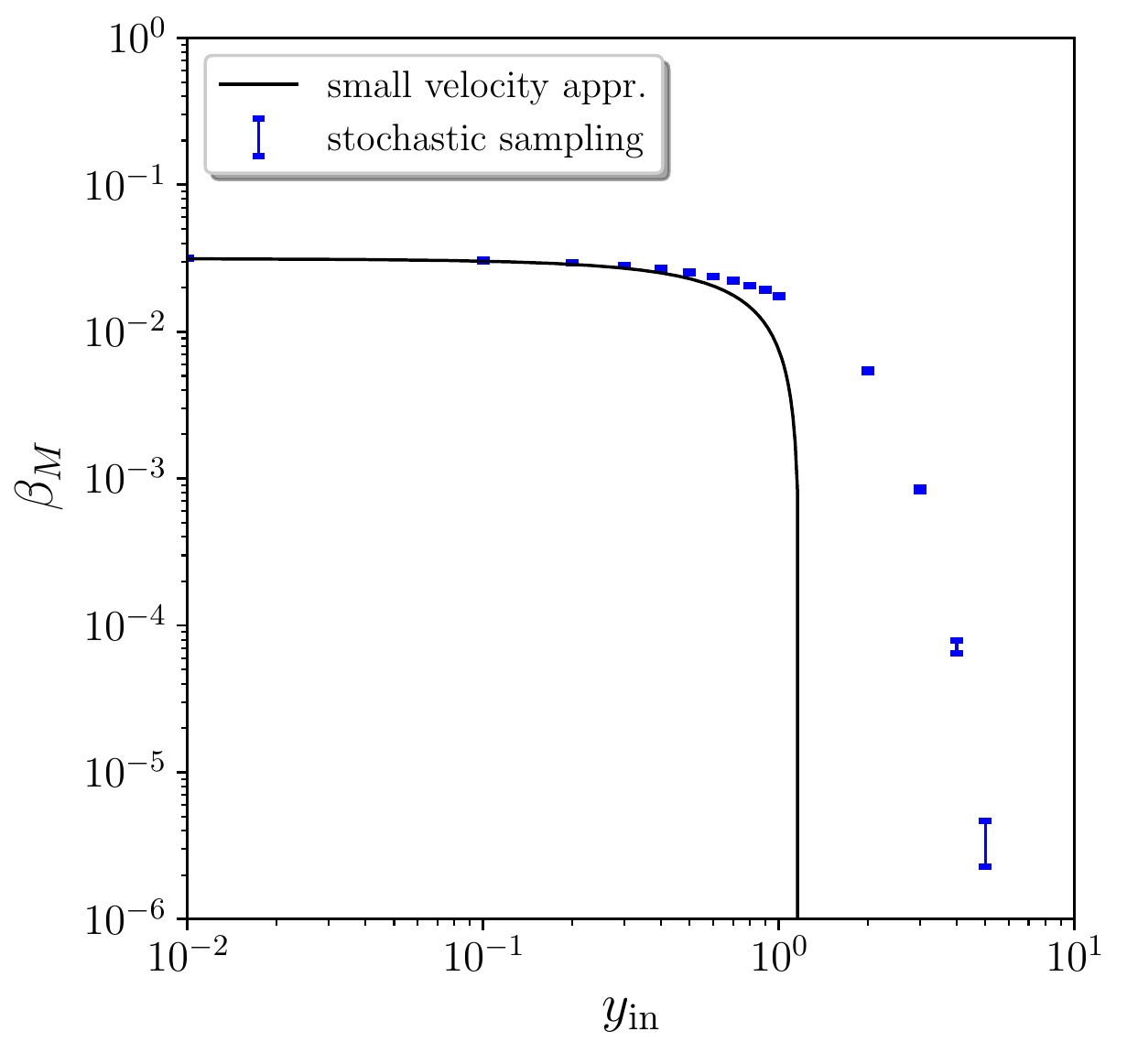}
    \caption{Fraction of the Langevin realisations that undergo a number of \efolds~$\mathcal{N}>\langle \mathcal{N} \rangle + \zeta_\uc$ with $\mu=1$ and $\zeta_\uc=1$. The black curves correspond to the small-$y$ approximation~\eqref{eq:beta:smally}, while the blue bars are reconstructed from a large number of numerical simulations of the Langvin equations \eqref{eq:langevin:xy} as in previous figures. \label{fig:beta:smally}}
\end{figure}

One can also calculate the mass fraction $\beta_M$ of primordial black holes from the estimate given in \Eq{eq:def:beta}, and which leads to
\bea \label{eq:beta:smally}
\beta_M
&= \sum_{n=0}^{\infty} \left\lbrace \frac{a_n^{(0)}(x,y)}{\Lambda_n^{(0)}} + \frac{a_n^{(1)}(x,y)}{\Lambda_n^{(0)}+3}\ee^{-3\left[\left<\mathcal{N}\right>(x,y)+\zeta_c\right]} \right\rbrace \ee^{-\Lambda_n^{(0)}\left[\left<\mathcal{N}\right>(x,y)+\zeta_\uc\right]} \, . 
\eea 
This expression is compared to our numerical results in \Fig{fig:beta:smally}, for $\zeta_\mathrm{c}=1$
When $y_\uin$ increases, $\beta_M$ decreases, and we must simulate a very large number of realisations in order to avoid being dominated by statistical noise (in practice one needs to produce many more realisations than $1/\beta_M$ in order to accurately compute $\beta_M$), which becomes numerically expensive. In practice, for $y_\mathrm{in} \geq 6$, we produced $10^8$ realisations and found that none of these experienced more than $\left< \mathcal{N} \right> + 1$ \efolds~of inflation, so we can only place an upper bound of $\beta_M< 10^{-8}$ for these values of the initial velocity. 

When $y_\uin$ is small, one can see that \Eq{eq:beta:smally} provides a good fit to the asymptotic behaviour of $\beta_M$ but is less accurate when $y_\uin$ approaches one (at least slightly less than the small-velocity approximation in \Figs{fig:meanefolds:smally} and~\ref{fig:pdf:smally}). This is because, as mentioned above, for values of $\mu$ of order one or below, the sets of eigenvalues highly overlap and the mass fraction picks up contribution from higher-order terms. The approximation becomes better when larger values of $\mu$ are employed. 

Let us also mention that the large-velocity approximation developed in \Sec{sec:classicallimit} vastly underestimates the mass fraction, since this expansion applies to the neighborhood of the maximum of the PDF~\cite{Pattison:2018bct, Ezquiaga:2019ftu}. For this reason we do not display it in \Fig{fig:beta:smally} since it lies many orders of magnitude below the actual result. 
\section{Volterra equation approach}
\label{sec:volterra}
The above considerations made clear that resolving the tail of the PDF in the presence of a substantial initial velocity is both analytically and numerically challenging. This is why in this section, we propose a new method, that makes use of Volterra integral equations, to solve the first-passage-time problem for the stochastic system depicted in \Fig{fig:sketch}. As we will see, it is much more efficient at reconstructing the tail of the PDF than by simulating a large number of Langevin realisations.

The first step of this approach is to make the dynamics trivial by introducing the phase-space variable 
\bea
z\equiv \frac{\mu}{\sqrt{2}}\left(x-y\right) ,
\eea
in terms of which \Eq{eq:langevin:xy} can be rewritten as
\bea
\label{eq:Langevin:z}
\frac{\dd z}{\dd N} = \xi\left(N\right) .
\eea
In the absence of any boundary conditions, starting from $z_\uin$ at time $N_\uin$, the distribution function of the $z$ variable at time $N$ is of the Gaussian form
\bea
\label{eq:sol:FP:noBound}
f\left(z,N\vert z_\uin,N_\uin\right) = \frac{\ee^{-\frac{\left(z-z_\uin\right)^2}{2\left(N-N_\uin\right)}}}{\sqrt{2\pi\left(N-N_\uin\right)}}\, .
\eea

The non-trivial features of the problem are now contained in the boundary conditions, since the absorbing boundary at $\phi=\phi_\uend$ ($x=0$) and the reflective boundary at $\phi=\phi_\uend+\Delta\phi_\mathrm{well}$ ($x=1$) give rise to the time-dependent absorbing and reflective boundaries, at respective locations
\bea
\label{eq:boundaries:z}
z_-(N) &= - \frac{\mu}{\sqrt{2}} y_\uin \ee^{-3\left(N-N_\uin\right)} \, ,\\
z_+(N) &= \frac{\mu}{\sqrt{2}}\left[1 - y_\uin \ee^{-3\left(N-N_\uin\right)}\right] .
\eea
In terms of the $z$ variable, these equations describe the motion of a free particle with no potential gradient and constant noise amplitude, within a well of fixed width but with moving boundaries, one being absorbing and the other one reflecting.
This problem has no known analytical solution but one can consider a related, solvable problem, where the two boundaries are absorbing. Let us first describe how that problem can be solved with Volterra equations, before explaining how the solutions to the original problem can be expressed in terms of solutions of the related one.
\subsection{Two absorbing boundaries}
\label{sec:two:absorbing:boundaries}

If the two boundaries are absorbing, following \Refa{Buonocore:1990vol}, one can introduce $\calP^{(0)}_-(N\vert N_\uin, z_\uin)$ [respectively  $\calP^{(0)}_+(N\vert N_\uin, z_\uin)$], the probability densities that $z$ crosses for the first time the boundary $z_-$ (respectively $z_+$) at time $N$ without having crossed $z_+$ (respectively $z_-$) before, starting from $z_\uin$ at time $N_\uin$, as well as the two auxiliary functions
\bea
\Psi_{\pm}\left(N\vert z_\uin,N_\uin\right)\equiv \left[z_\pm'(N)-\frac{z_\pm(N)-z_\uin}{N-N_\uin}\right] f\left[z_\pm(N),N\vert z_\uin,N_\uin\right] .
\eea
One can then show that $\calP^{(0)}_-(N\vert N_\uin, z_\uin)$ and $\calP^{(0)}_+(N\vert N_\uin, z_\uin)$ satisfy the two coupled integral equations~\cite{Buonocore:1990vol}
\bea
\label{eq:Volterra}
\calP^{(0)}_-(N\vert z_\uin,N_\uin) = & \Psi_-\left(N\vert z_\uin,N_\uin\right)-\int_{N_\uin}^N \dd \tilde{N} \Big\lbrace \calP^{(0)}_-(\tilde{N}\vert  z_\uin, N_\uin) \Psi_-\left[N\vert z_-(\tilde{N}),\tilde{N}\right]
 \\ &
+\calP^{(0)}_+(\tilde{N}\vert  z_\uin, N_\uin) \Psi_-\left[N\vert z_+(\tilde{N}),\tilde{N}\right]\Big\rbrace\, ,\\
\calP^{(0)}_+(N\vert  z_\uin, N_\uin) = & -\Psi_+\left(N\vert z_\uin,N_\uin\right)+\int_{N_\uin}^N \dd \tilde{N} \Big\lbrace \calP^{(0)}_-(\tilde{N}\vert z_\uin, N_\uin) \Psi_+\left[N\vert z_-(\tilde{N}),\tilde{N}\right]
 \\ &
+\calP^{(0)}_+(\tilde{N}\vert  z_\uin, N_\uin) \Psi_+\left[N\vert z_+(\tilde{N}),\tilde{N}\right]\Big\rbrace\, .
\eea
These formulas are obtained from first principles in \Refa{Buonocore:1990vol} and we do not reproduce their derivation here. It is nonetheless worth pointing out that they can be generalised to any drift function and noise amplitude in the Langevin equation, although, for practical use, one needs to compute the solution $f$ of the Fokker-Planck equation in the absence of boundaries [as was done in \Eq{eq:sol:FP:noBound} here], which is not always analytically possible. These equations can be readily solved numerically by discretising the integrals. Starting from $\calP^{(0)}_-(N_\uin\vert N_\uin, z_\uin)=\calP^{(0)}_+(N_\uin\vert N_\uin, z_\uin)=0$ (if $z_\uin \in [z_-(N_\uin),z_+(N_\uin)]$), one can then compute the value of the distribution functions at each time step by plugging into the right-hand sides of \Eqs{eq:Volterra} their values at previous time steps. 
\subsection{One absorbing and one reflective boundary}
Let us now consider again our original problem, where the boundary at $z_-(N)$ is absorbing and the boundary at $z_+(N)$ is reflective, and where our goal is to compute the probability density $\calP_-(N\vert z_\uin, N_\uin)$ that the system crosses the absorbing boundary at time $N$ starting from $z_\uin$ at time $N_\uin$.

This can be done by considering the probability density that the system crosses the absorbing boundary at time $N$ after having bounced exactly $n$ times at the reflective boundary, and that we denote $\calP^{(n)}_-(N\vert z_\uin,N_\uin)$. One can decompose 
\bea
\label{eq:PDF:sum:reflectionNumber}
\calP_-(N\vert z_\uin,N_\uin)  = \sum_{n=0}^\infty \calP^{(n)}_-(N\vert z_\uin, N_\uin)\, .
\eea
We also introduce $\calP^{(n)}_+(N\vert z_\uin, N_\uin)$, which is the distribution function associated with the $(n+1)^{\mathrm{th}}$ bouncing time. Obviously, neither $\calP^{(n)}_-$ or $\calP^{(n)}_+$ are normalised to unity since all realisations of the stochastic process escape through the absorbing boundary after a finite number of bouncing events. More precisely, their norms decay with $n$. In the regimes where this makes the sum~\eqref{eq:PDF:sum:reflectionNumber} converge, this can be used to compute $\calP_-$ using the following iterative procedure.

For $n=0$, it is clear that $\calP^{(0)}_\pm$ match the quantities denoted in the same way in \Sec{sec:two:absorbing:boundaries} (hence the notation), since $\calP^{(0)}_-$ corresponds to the probability that the system crosses the absorbing boundary without having ever bounced at the reflective boundary, and $\calP^{(0)}_+$ is associated with the first encounter with the reflective boundary. 

For $n\geq 1$, let us consider realisations that bounce $n$ times before exiting the well, and let us call $N_\mathrm{b}$ their last bouncing time (\ie their $n^\mathrm{th}$ bouncing time). The distribution function associated with $N_\mathrm{b}$ is simply $\calP^{(n-1)}_+(N\vert z_\uin, N_\uin)$. Then, starting from the reflective boundary at time $N_\mathrm{b}$, one has to determine the probability that the system crosses the absorbing boundary at time $N$ without bouncing again against the reflective boundary. From the above definitions, this probability is nothing but $\calP_-^{(0)}[N\vert z_+(N_{\mathrm{b}}),N_{\mathrm{b}}]$. This gives rise to 
 \bea
 \calP^{(n)}_-(N\vert z_\uin,N_\uin) &=& \int_{N_\uin}^N \dd N_{\mathrm{b}}  \calP^{(n-1)}_+(N_\mathrm{b}\vert z_\uin, N_\uin) \calP_{-}^{(0)}\left[N\vert z_+(N_{\mathrm{b}}),N_{\mathrm{b}}\right].
 \eea
 Similarly, once at the reflective boundary for the $n^{\mathrm{th}}$ time at time $N_\mathrm{b}$, the probability to hit the reflective boundary for the $(n+1)^{\mathrm{th}}$ time at time $N$, without exiting through the absorbing boundary before then, is given by $\calP_+^{(0)}[N\vert z_+(N_\mathrm{b}),N_\mathrm{b}]$, which leads to
 \bea
 \label{eq:Pplus_n_Volterra}
  \calP^{(n)}_+(N\vert z_\uin, N_\uin) &=& \int_{N_\uin}^N \dd N_{\mathrm{b}}  \calP^{(n-1)}_+(N_\mathrm{b}\vert z_\uin, N_\uin) \calP_{+}^{(0)}\left[ N\vert z_+(N_{\mathrm{b}}),N_{\mathrm{b}}\right] .
 \eea
These allow one to compute the distribution functions $ \calP^{(n)}_\pm$ iteratively, hence to obtain the full first-exit-time distribution function from \Eq{eq:PDF:sum:reflectionNumber}. 

\begin{figure} 
    \centering
    \includegraphics[width=0.69\textwidth]{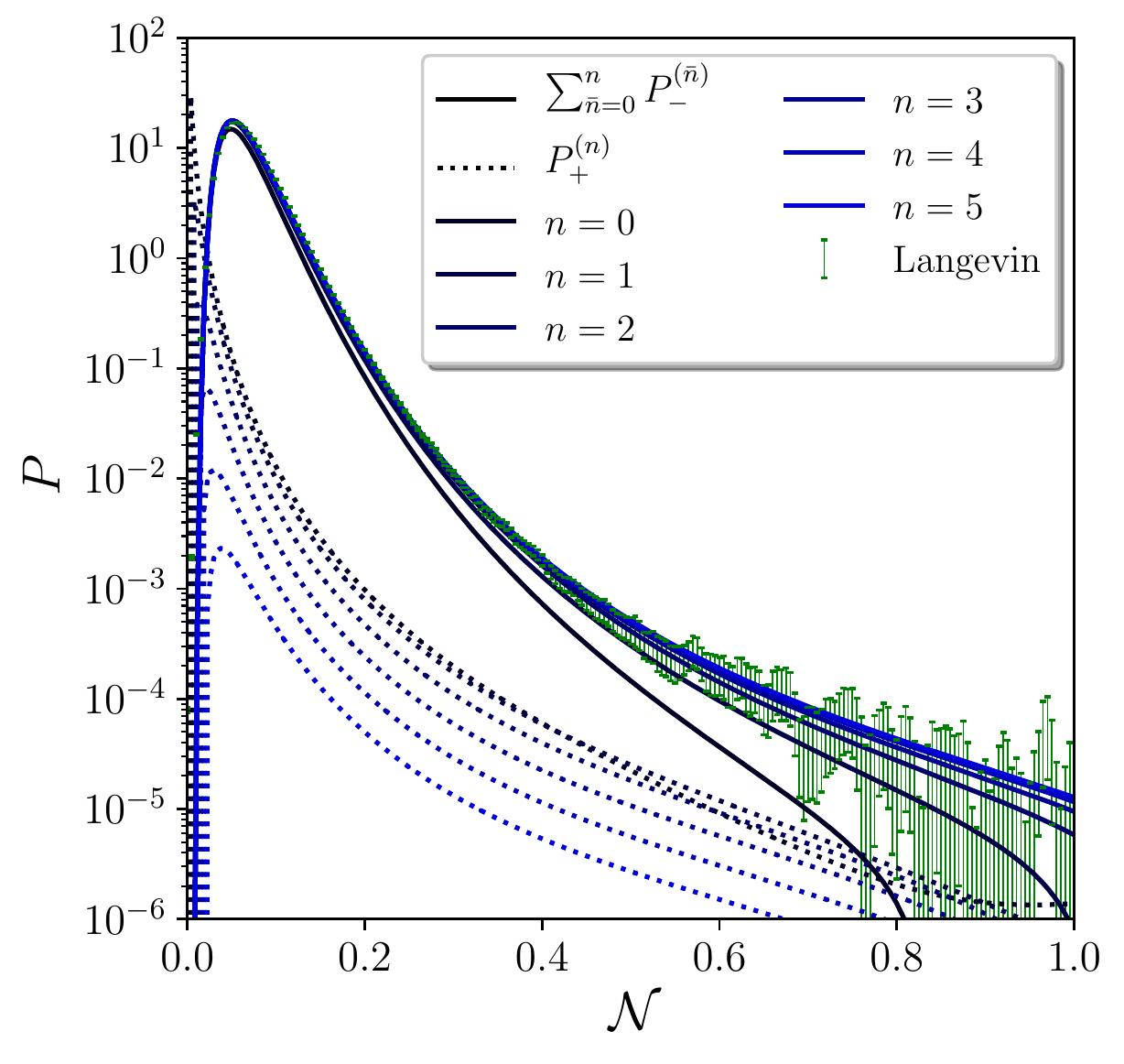}
    \caption{Distribution function of the first passage time $\mathcal{N}$ for $\mu=1$ and $y_\uin=5$, computed from the Volterra integral equations~\eqref{eq:Volterra}. The solid lines correspond to \Eq{eq:PDF:sum:reflectionNumber} truncated at various order $n$ labeled with different colours. One can check that, as $n$ increases, this approaches the distribution function reconstructed from direct simulations of the Langevin equations and displayed in green.
    The dashed lines stand for $\calP_+^{(n)}$ given in \Eq{eq:Pplus_n_Volterra}, which correspond to the distribution function of the $(n+1)^\mathrm{th}$ bouncing time.     \label{fig:Volterra}}
\end{figure}

For illustrative purpose, this is done in \Fig{fig:Volterra} for $\mu=1$ and $y_\uin=5$. As $n$ increases, one can check that the distribution functions of the $(n+1)^\mathrm{th}$ bouncing time, $\calP_+^{(n)}$, decay (dotted lines). As a consequence, when truncating \Eq{eq:PDF:sum:reflectionNumber} at increasing orders (solid lines), one quickly approaches the distribution reconstructed from Langevin simulations (green bars). Close to the maximum of the distribution, the result from the first iterations already provide an excellent approximation, while on the tail, one has to iterate a few times before reaching a good agreement. There, the statistical error associated to the Langevin reconstruction is large since the statistics is sparse, and the Volterra technique is particularly useful.

In practice, one has to truncate the sum \eqref{eq:PDF:sum:reflectionNumber} at a fixed order $n_\umax$. As already argued, the convergence of the sum is improved in situations where $ \calP^{(n)}_\pm$ quickly decays with $n$, \ie when it is unlikely to undergo a large number of bouncing events. This implies that this computational scheme is well suited in the ``drift-dominated'' (or large initial velocity) regime of \Sec{sec:classicallimit}, and we find small values of $y_\uin$ to be indeed numerically more challenging.
\subsection{Revisiting the drift-dominated limit}
The Volterra equations~\eqref{eq:Volterra} also enable to expand the first-passage-time distribution functions around any known limiting case, by iteratively plugging the approximated distributions on the right-hand sides of \Eqs{eq:Volterra} and reading their improved version on the left-hand sides.

One such regime is the drift-dominated limit, or large-velocity limit, already studied in \Sec{sec:classicallimit}. In this limit, at leading order, each realisation of the stochastic process follows the classical path, hence
\bea
\label{eq:pdf:cl:lo:volterra}
\calP_-^{\mathrm{cl,LO}}\left(N\vert z_\uin,N_\uin\right) = \delta\left[N-N_\ucl(z_\uin,N_\uin)\right],
\eea
see \Eq{eq:pdf:cl:lo}, where the classical number of \efolds~is simply obtained by requiring that $z_-(N_\ucl)=z_\uin$ (since $z$ remains still in the absence of quantum diffusion), which from \Eq{eq:boundaries:z} gives rise to 
\bea
N_\ucl(z_\uin,N_\uin) = N_\uin - \frac{1}{3}\ln\left(-\frac{z_\uin}{y_\uin}\frac{\sqrt{2}}{\mu}\right),
\eea
in agreement with \Eq{eq:meanefolds:classical:LO}. 

Along the classical path, the reflective boundary plays no role, hence whether it is absorbing or reflective is irrelevant and \Eq{eq:pdf:cl:lo:volterra} can also serve as the leading-order expression for $\calP_-^{(0)}$. At that order, one simply has $\calP_+^{(0)}(N\vert z_\uin, N_\uin)=0$. Plugging these formulas into the right-hand sides of \Eqs{eq:Volterra}, one obtains, at next-to-leading order,
\bea
\label{eq:P(0):cl:NLO}
\calP^{(0),\mathrm{cl,NLO}}_-(N\vert N_\uin, z_\uin) =&
\left[z_-'(N)-\frac{z_-(N)-z_\uin}{N-N_\uin}\right]\frac{\ee^{-\frac{\left[z_-(N)-z_\uin\right]^2}{2(N-N_\uin)}}}{\sqrt{2\pi(N-N_\uin)}}
\\ & \kern-8em
-\left[
z_-'(N)-\frac{z_-(N)- z_\uin}{N-N_\ucl(z_\uin,N_\uin)}\right]
\frac{\ee^{-\frac{\left[
z_-(N)-z_\uin\right] ^2}{2 \left[N-N_\ucl(z_\uin,N_\uin)\right]}}}{\sqrt{2\pi\left[N-N_\ucl(z_\uin,N_\uin)\right]}}\theta\left[N-N_\ucl(z_\uin,N_\uin)\right]\\
\calP^{(0),\mathrm{cl,NLO}}_+(N\vert N_\uin, z_\uin) =&
-\left[z_+'(N)-\frac{z_+(N)-z_\uin}{N-N_\uin}\right]\frac{\ee^{-\frac{\left[z_+(N)-z_\uin\right]^2}{2(N-N_\uin)}}}{\sqrt{2\pi(N-N_\uin)}}
\\ & \kern-8em
+\left[ 
z_+'(N)-\frac{z_+(N)- z_\uin}{N-N_\ucl(z_\uin,N_\uin)}\right]
\frac{\ee^{-\frac{\left[
z_+(N)-z_\uin\right] ^2}{2 \left[N-N_\ucl(z_\uin,N_\uin)\right]}}}{\sqrt{2\pi\left[N-N_\ucl(z_\uin,N_\uin)\right]}}\theta\left[N-N_\ucl(z_\uin,N_\uin)\right].
\eea
In order to gain further analytical insight, one can expand $\calP^{(0)}_-$ close to $N=N_\ucl$, where it is expected to peak. One obtains
\bea
\calP^{(0),\mathrm{cl,NLO}}_-(N\vert N_\uin, z_\uin) \simeq & \frac{-3z_\uin}{\sqrt{2\pi \left[N_\ucl(z_\uin,N_\uin)-N_\uin\right]}}\ee^{-\frac{9 z_\uin^2}{2}\frac{\left[N-N_\ucl(z_\uin,N_\uin)\right]^2}{N_\ucl(z_\uin,N_\uin)}}\, ,
\eea
which is nothing but a Gaussian distribution centred on $N=N_\ucl$ and with variance $\langle \delta N^2\rangle = (N_\ucl-N_\uin)/(9  z_\uin^2)$, in agreement with \Eq{eq:deltaN2:cl:nlo} [however, here, we are able to reconstruct the Gaussian PDF, something that was not directly possible with the expansion of \Sec{sec:classicallimit}, see the discussion around \Eq{eq:fnl}].

From \Eq{eq:P(0):cl:NLO}, one can see that the exponential terms in $\calP^{(0)}_+$ rather peak at the time $N_+$ such that $z_+(N_+)=z_\uin$, namely
$
N_+ = N_\uin-\frac{1}{3}\ln\left(\frac{1}{y_\uin}-\frac{z_\uin}{y_\uin}\frac{\sqrt{2}}{\mu}\right)=
N_\uin-\frac{1}{3}\ln\left(1+\frac{1-x_\uin}{y_\uin}\right).
$
If one starts at the location of the reflective boundary, $x_\uin=1$, then one simply has $N_+=N_\uin$ as one may have expected, but otherwise $N_+<N_\uin$. This means that $\calP^{(0)}_+$ is maximal near the origin $N=N_\uin$, and can be expanded around there,
\bea
\calP^{(0),\mathrm{cl,NLO}}_+(N\vert N_\uin, z_\uin) \simeq &
\left[\frac{\mu\left(1-x_\uin\right)}{2\sqrt{\pi}\left( N-N_\uin\right)^{3/2}}
+\frac{9 y_\uin \mu}{4}\sqrt{\frac{N-N_\uin}{\pi}}
\right]
\\ & \kern-5em
\exp
\left[
-\frac{\left(1-x_\uin\right)^2 \mu^2}{4\left(N-N_\uin\right)}+
\frac{3}{2}\mu^2 y_\uin\left(x_\uin-1\right)-\frac{9}{4}\mu^2 y_\uin \left(x_\uin + y_\uin -1 \right)\left(N-N_\uin\right)\right],
\eea
where the expansion has been performed at next-to-leading order in $(N-N_\uin)$ such as to also describe the case $x_\uin=1$ for which the leading order result vanishes. One can use this approximation to integrate $\calP_+^{(0)}$ and estimate the probability $p_+$ that the system bounces at least once on the reflective boundary,
\bea
p_+\simeq
\begin{cases}
\ee^{-\frac{3}{2}y_\uin \mu^2 \left(1-x_\uin\right)}\quad\mathrm{if}\quad x_\uin<1\\
\frac{1}{3\mu^2 y_\uin^2}\quad\mathrm{if}\quad x_\uin=1
\end{cases}
.
\eea
One can check that, for large initial velocities, the suppression is much less pronounced in the case where $x_\uin=1$: indeed, if one starts precisely at the location of the reflective boundary, the probability to ``bounce'' against it (although this becomes a somewhat singular quantity) is larger than if one starts away from it. One can check that, when $\mu y_\uin \gg 1$, this probability is small, in agreement with the results of \Sec{sec:classicallimit} where it was shown that $(\mu y_\uin)^{-1} $ is indeed the relevant parameter to perform the drift-dominated expansion with. 

It is also interesting to notice that already at next-to-leading order, the distributions~\eqref{eq:P(0):cl:NLO} are heavy-tailed. Indeed, when $N\gg N_\ucl(z_\uin, N_\uin)$, one has $z_- \simeq z_-' \simeq z_+'\simeq 0$ and $z_+\simeq \sqrt{\mu/2}$, which gives rise to
\bea
\calP^{(0),\mathrm{cl,NLO}}_- &\simeq 3 z_\uin \frac{N_\uin-N_\ucl}{2 \sqrt{2\pi}} N^{-5/2}\\
\calP^{(0),\mathrm{cl,NLO}}_+ &\simeq \mu \frac{N^{-3/2}}{2\sqrt{\pi}} \, .
\eea
These power-law behaviours, which imply that not all moments of the distributions are defined, typically arise in situations where a single absorbing boundary condition is imposed and the inflating field domain is unbounded, a typical example being the Levy distribution. At this order, indeed, $\calP^{(0),\mathrm{cl,NLO}}_-$ still carries no information about the reflective boundary [since $z_+$ does not appear explicitly in the first of \Eqs{eq:P(0):cl:NLO}]. This is no-longer the case at next-to-next-to-leading order, although unfortunately, there, the integrals appearing in \Eq{eq:Volterra} can no longer be performed analytically.
\section{Applications}
\label{sec:applications}
So far we have studied stochastic effects in USR inflation by focusing on the toy model depicted in \Fig{fig:sketch}, where the inflationary potential contains a region that is exactly flat. In this section, we want to consider more generic potentials, and see how the toy model we have investigated helps to describe more realistic setups.
\subsection{The Starobinsky model}
\label{sec:StarobinskyModel}
\begin{figure}
    \centering
    \includegraphics[width=0.7\textwidth]{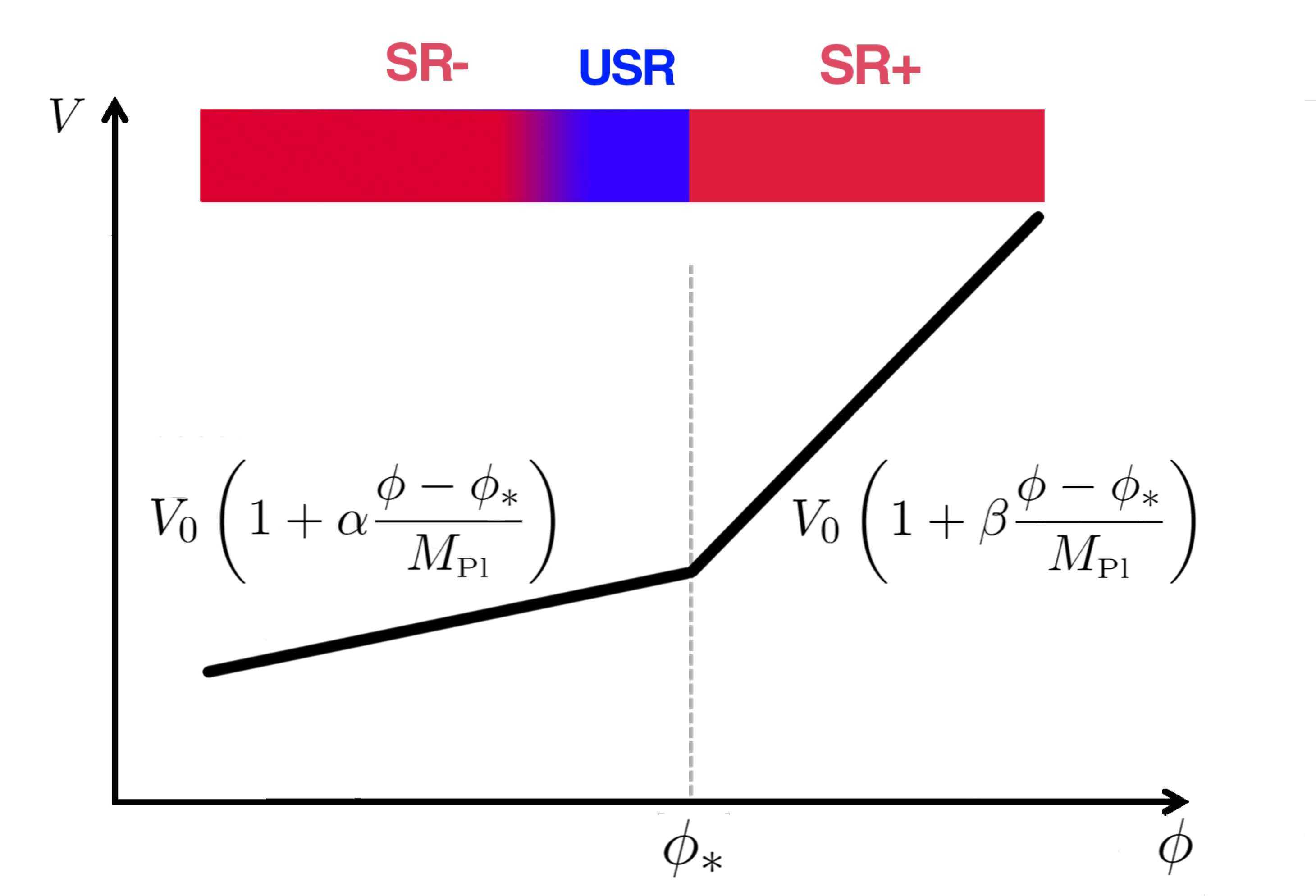}
    \caption{A sketch of the piece-wise linear potential~\eqref{eq:Staro:potential}, known as the Starobinsky model. When reaching the second branch of the potential where the potential slope suddenly drops, the inflaton experiences a phase of USR inflation before progressively relaxing back to slow roll. \label{fig:staro:potential}}
\end{figure}

Let us start by considering the Starobinsky potential for inflation, where the potential has two linear segments with different gradients, namely
\bea 
\label{eq:Staro:potential}
V(\phi) = \begin{cases}
\displaystyle
V_{0}\left(1+\alpha\frac{\phi-\phi_*}{\Mp}\right) &\quad \mathrm{for}\quad  \phi< \phi_* \\ 
& \\
\displaystyle
V_{0}\left(1+\beta\frac{\phi-\phi_*}{\Mp}\right) &\quad \mathrm{for}\quad \phi> \phi_* 
\end{cases}
\, ,
\eea
where $\beta>\alpha>0$, see \Fig{fig:staro:potential}.
For $\beta\ll 1$, we may take $H$ to be approximately constant throughout, and this model begins with a first phase of slow-roll inflation when $\phi>\phi_*$. This is followed by a phase of USR inflation for $\phi<\phi_*$ caused by the discontinuous jump in the gradient of the potential kicking the inflaton off the slow-roll trajectory at $\phi=\phi_*$.
Finally, since USR is unstable in this model (note that it can be stable in other setups~\cite{Pattison:2018bct}), the inflaton relaxes back to the slow-roll regime in a second phase of slow-roll inflation. 

The velocity at the beginning of the USR phase is given by the slow-roll attractor before the discontinuity, namely
\bea \label{eq:initalvelocity:staro}
\dot{\phi}_* \simeq -\Mp H \beta \, .
\eea 
In the absence of quantum diffusion, the width of the USR well can be obtained by finding the point at which the inflaton relaxes back to slow roll. Taking $H$ to be constant, the Klein-Gordon equation~\eqref{eq:kleingordon} in the second branch of the potential can be solved to give
\bea
\label{eq:Staro:model:phi(t)}
\frac{\phi(t)-\phi_* }{\Mp}=  - \alpha H\left(t-t_*\right)+\frac{\beta-\alpha}{3}\left[\ee^{-3H(t-t_*)}-1\right],
\eea 
where we have made use of the initial condition~\eqref{eq:initalvelocity:staro} at $\phi=\phi_*$. The velocity of the inflaton, $\dot{\phi}$, is then
\bea
\label{eq:Staro:momentum}
\frac{\dot{\phi}(t)}{\Mp H } = -\alpha - \left(\beta-\alpha\right) \ee^{-3H(t-t_*)}.
\eea 
The first term, $-\alpha$, represents the slow-roll attractor towards which the system asymptotes at late time, while the second term is the USR velocity. It provides the main contribution to the total velocity at the onset of the second branch of the potential if $\beta\gg\alpha$, and that condition ensures that the inflaton experiences a genuine USR phase. The two contributions are equal at the time $t_{\mathrm{USR}\to\mathrm{SR}} -t_*=\ln(\beta/\alpha-1)/(3H)$, and by evaluating \Eq{eq:Staro:model:phi(t)} at that time, one finds~\cite{Pattison:2019hef}
\bea 
\frac{\phi_{\mathrm{USR}\to\mathrm{SR}}-\phi_*}{\Mp} = -\frac{\alpha}{3}\ln\left(\frac{\beta-\alpha}{\alpha}\right)+\frac{2\alpha-\beta}{3}\, .
\eea 
This allows one to evaluate the width of the USR well, $\Delta\phi_\mathrm{well}$, and the critical velocity with \Eq{eq:critical:velocity}. Since the initial velocity corresponds to \Eq{eq:initalvelocity:staro}, one obtains for the rescaled initial velocity
\bea
y_\uin = \frac{1}{1+\frac{\alpha}{\beta} \left[\ln\left(\frac{\beta-\alpha}{\alpha}\right)-2\right]} \simeq 1
\quad\text{if} \quad\beta\gg \alpha\, .
\eea 
The fact that $y_\uin \simeq 1$ in this model should not come as a surprise, since the width of the USR well was defined as being where the classical trajectory relaxes to slow-roll, so by definition, the initial velocity of the inflaton is precisely the one such that the field reaches that point in finite time. The Starobinsky model therefore lies at the boundary between the drift-dominated and the diffusion-dominated limits studied in \Secs{sec:classicallimit} and~\ref{sec:stochasticlimit} respectively. 

The value of the parameter $\mu$ can be read of from \Eq{eq:def:mu} and is given by
\bea
\label{eq:mu:Staro:TM}
\mu\simeq \frac{2\sqrt{2}}{3}\pi \beta \frac{\Mp}{H}\, .
\eea 
Recall that in order for the potential to support a phase of slow-roll inflation in the first branch, we have taken $\beta\ll 1$. 
However, for inflation to proceed at sub-Planckian energies, $H/\Mp\ll 1$, and the current upper bound on the tensor to scalar ratio~\cite{Akrami:2018odb} imposes $H/\Mp\lesssim 10^{-5}$ in single-field slow-roll models. Therefore, unless $\beta<10^{-5}$, one has $\mu\gg 1$, and as argued below \Eq{eq:usr:efolds:2ndmoment:analytic}, $y\sim 1$ corresponds to where an abrupt transition between the diffusion-dominated and the drift-dominated regime takes place. Hence neither of the approximations developed in \Secs{sec:classicallimit} and~\ref{sec:stochasticlimit} really apply, and it seems difficult to predict the details of the PBH abundance without performing numerical explorations of the model. If $\beta\ll H/\Mp$, then $\mu\ll 1$ and $y\sim 1$ lies on the edge of the diffusion-dominated regime where the abundance of PBHs is highly suppressed in that case.

Let us note however that the above description is not fully accurate since we have used the classical solution~\eqref{eq:Staro:model:phi(t)} to estimate when and where the transition between USR and the last slow-roll phase takes place. While this may be justified in the drift-dominated regime, we have found that $y_\uin \simeq 1$ so this approximation is not fully satisfactory in the presence of quantum diffusion. In practice, different realisations may exit the USR phase at different values of $\phi_{\mathrm{USR}\to\mathrm{SR}}$, and we may not be able to map this problem to the toy model discussed in \Secs{sec:classicallimit} and~\ref{sec:stochasticlimit}. In fact, since quantum diffusion does not affect the dynamics of the momentum in USR, see \Eq{eq:eom:v:stochastic}, \Eq{eq:Staro:momentum} remains correct even in the presence of quantum diffusion, \ie $\bar{\pi}/\Mp = -\alpha -(\beta-\alpha)\ee^{-3(N-N_*)} $, where we still assume that $H$ is almost constant. As a consequence, the transition towards the slow-roll regime occurs at the same time (rather than at the same field value) for all realisations, namely~\cite{Pattison:2019hef} 
\bea
\mathcal{N}_{\mathrm{USR}\to\mathrm{SR}} = \frac{1}{3}\ln\left(\frac{\beta}{\alpha}-1\right)\, ,
\eea 
which is obtained by equating the two terms in the above expression for $\bar{\pi}/\Mp$.
\begin{figure} 
    \centering
    \includegraphics[width=0.515\textwidth]{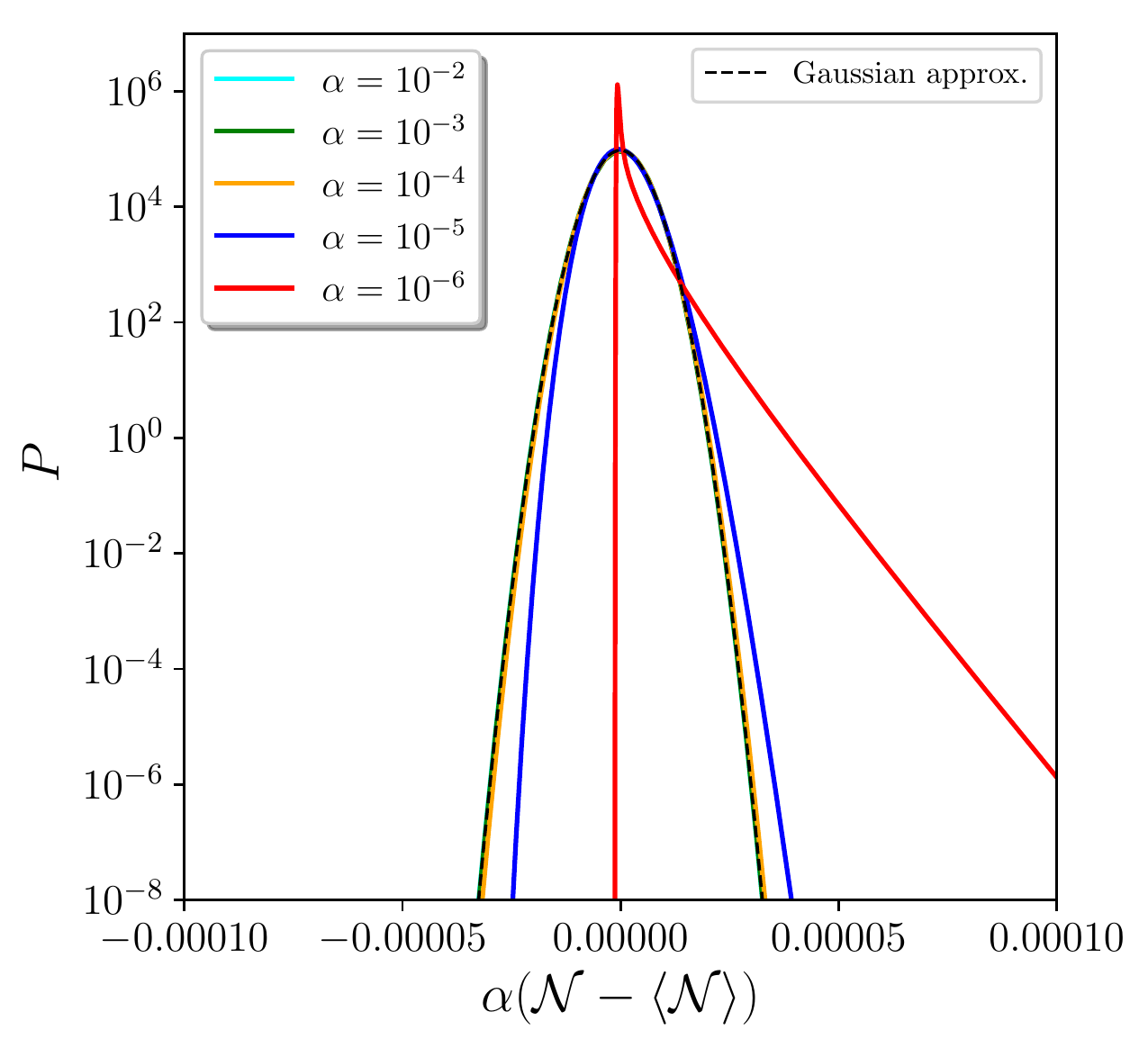}
    \caption{Distribution function of the number of \efolds~realised in the Starobinsky model~\eqref{eq:Staro:potential}, between $\phi_*$ and $\phi_\uend$, obtained from the iterative Volterra procedure detailed in \Sec{sec:volterra} and extended in \Eqs{eq:z+:z_:Starobinsky}. We have used $H/\Mp=10^{-5}$ and $\beta=0.1$ while letting $\alpha$ vary, and we have set $\phi_\uend$ such that the classical trajectory~\eqref{eq:Staro:model:phi(t)} realises $7$ \efolds~between $\phi_*$ and $\phi_\uend$. The curves for $\alpha=10^{-2}$, $\alpha=10^{-3}$ and $\alpha=10^{-4}$ are so close to each other that they cannot be distinguished by eye.
 \label{fig:Starobinsky:potential:P(N)}}
\end{figure}
The problem should therefore be reconsidered, and conveniently, the methods developed in \Sec{sec:volterra} can be extended to the case of a linear potential. Indeed, one can introduce the new field variable
\bea
z &= \frac{\bar{\phi}+\frac{\bar{\pi}(N)}{3}+\Mp\alpha(N-N_*)}{H/(2\pi)}\\
&=2\pi\frac{\Mp}{H}\left[\frac{\bar{\phi}}{\Mp}+\alpha\left(N-N_*-\frac{1}{3}\right)+\frac{\alpha-\beta}{3}\ee^{-3(N-N_*)}\right]
\eea 
and, from \Eqs{eq:eom:phi:stochastic} and~\eqref{eq:eom:v:stochastic}, show that it undergoes free diffusion and obeys \Eq{eq:Langevin:z}. Assuming, without loss of generality, that inflation ends at $\bar{\phi}=0$, the location of the absorbing boundary, $z_-(N)$, and of the reflective boundary, $z_+(N)$, respectively correspond to $\bar{\phi}_-=0$ and $\bar{\phi}_+=\phi_*$, and are thus given by
\bea
\label{eq:z+:z_:Starobinsky}
z_-(N) & = 2\pi\frac{\Mp}{H}\left[\alpha\left(N-N_*-\frac{1}{3}\right)+\frac{\alpha-\beta}{3}\ee^{-3(N-N_*)}\right]\\
z_+(N) & = 2\pi\frac{\Mp}{H}\left[\frac{\phi_*}{\Mp}+\alpha\left(N-N_*-\frac{1}{3}\right)+\frac{\alpha-\beta}{3}\ee^{-3(N-N_*)}\right]\, .
\eea 
By plugging these formulas in the iterative Volterra procedure outlined in \Sec{sec:volterra}, one can then extract the PDF of the number of \efolds~without performing any approximation. The result is displayed in \Fig{fig:Starobinsky:potential:P(N)} for $H/\Mp=10^{-5}$ and $\beta=0.1$, and for a few values of $\alpha$. 

At leading order in quantum diffusion, one can use the classical $\delta N$ formalism to assess $\delta N \simeq \delta \phi / (\partial\phi/\partial N) = H/(2\pi)/ (\partial\phi/\partial N)$, where $\partial\phi/\partial N$ needs to be evaluated with the classical trajectory~\eqref{eq:Staro:model:phi(t)} (and given that the noise in the momentum direction is negligible both in the USR and slow-roll phases, see the discussion in \Sec{sec:StochasticInflation}). The integrated variance in the number of \efolds, $\langle \delta \mathcal{N}^2\rangle = \int (\delta N)^2 \dd N_\ucl $, can then be worked out, and in the limit where $\beta\gg \alpha$ and $N_\ucl\gg \mathcal{N}_{\mathrm{USR}\to\mathrm{SR}} $, one obtains
\bea
\label{eq:StarobinskyModel:Gaussian:appr}
\sqrt{\langle \delta\mathcal{N}^2 \rangle_\ucl}\simeq \frac{H}{2\pi} \frac{\sqrt{N_\ucl}}{\alpha\Mp}. 
\eea 
In this regime, one therefore recovers the same result as in slow-roll inflation, and the USR phase only introduces corrections that are suppressed by $\alpha/\beta$ and by $\ee^{3(\mathcal{N}_{\mathrm{USR}\to\mathrm{SR}}-N_\ucl)}$ (the full expression can readily be obtained but we do not reproduce it here since it is not particularly illuminating). The fact that $\delta \mathcal{N}$ scales as $1/\alpha$ in this limit is the reason why the distribution of $\alpha \delta \mathcal{N}$ is displayed in \Fig{fig:Starobinsky:potential:P(N)}, where the black dashed line stands for a Gaussian distribution with standard deviation given by \Eq{eq:StarobinskyModel:Gaussian:appr}. 

One can see that when $\alpha\geq 10^{-4}$, it provides an excellent fit to the full result, which therefore simply follows the classical Gaussian profile. In these cases, the PBH mass fraction $\beta_M$, given by \Eq{eq:def:beta}, is such that $\beta_M<10^{-100}$, where this upper bound simply corresponds to the numerical accuracy of our code (and where we use $\zeta_\uc=1$). When $\alpha=10^{-5}$ (blue curve), the distribution deviates from the classical Gaussian profile away from the immediate neighbourhood of its maximum: it gets skewed and acquires a heavier tail. In that case we find $\beta_M\simeq 0.0204$ while the classical Gaussian profile would give $\beta_M\simeq 0.0176$, hence it would slightly underestimate the amount of PBHs. When $\alpha=10^{-6}$ (red curve), the distribution is highly non Gaussian, and clearly features an exponential tail. We find $\beta_M\simeq 0.231$ in that case, while the classical Gaussian profile would give $\beta_M\simeq 0.81$, hence it would slightly underestimate the amount of PBHs. This is because, although the PDF has an exponential, heavier tail, it is also more peaked around its maximum. In those last two cases, PBHs are therefore overproduced (unless they correspond to masses that Hawking evaporate before big-bang nucleosynthesis, in which case they cannot be directly constrained, see \Refa{Papanikolaou:2020qtd}).

Notice that the value of $\alpha$ at which the result starts deviating from the classical Gaussian profile corresponds to $\alpha \sim H/\Mp$, which is what is expected for slow-roll inflation in a potential that is linearly tilted between $\phi_{\mathrm{USR}\to\mathrm{SR}}$ and $\phi_\uend$, see \Refa{Ezquiaga:2019ftu} [this is also consistent with \Eq{eq:StarobinskyModel:Gaussian:appr}]. As a consequence, the heavy tails obtained for small values of $\alpha$ are provided by the slow-roll phase that follows USR, and not so much by the USR phase itself. We conclude that the existence of a phase of USR inflation does not lead, per se, to substantial PBH production in the Starobinsky model.  

\subsection{Potential with an inflection point}
\label{sec:InflectionPoint:Plateau}
\begin{figure} 
    \centering
    \includegraphics[width=0.515\textwidth]{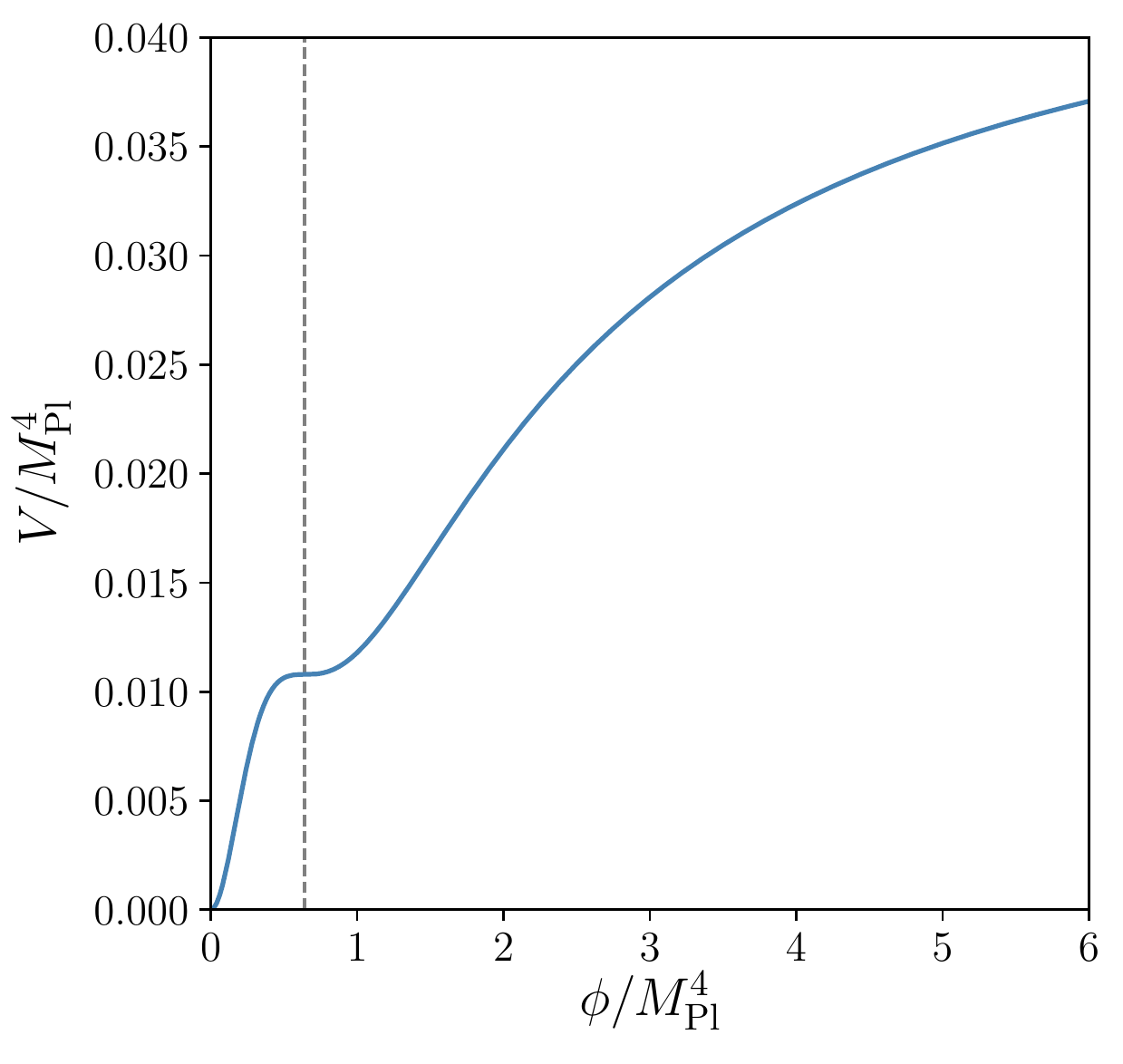}
    \includegraphics[width=0.47\textwidth]{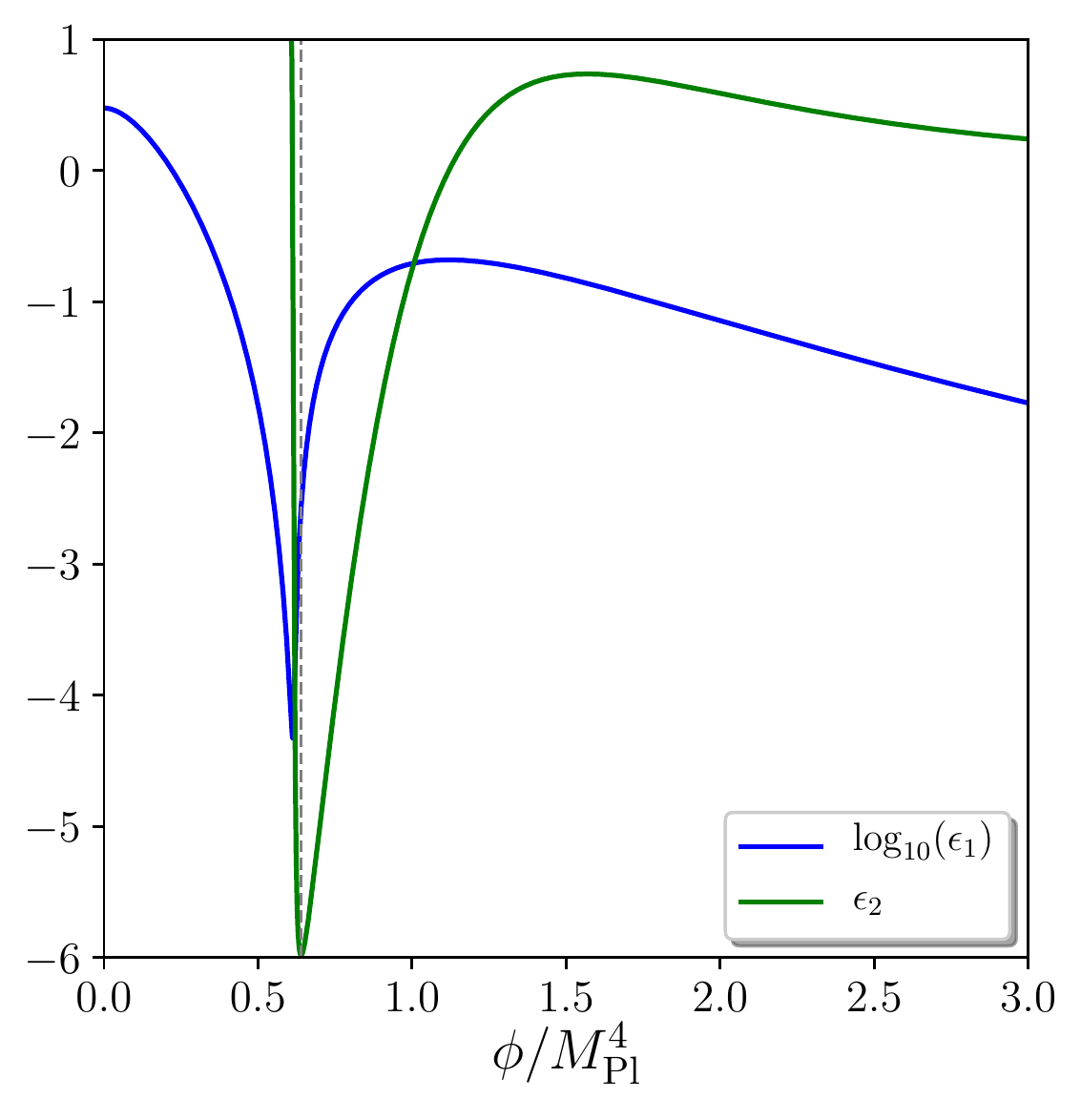}
    \caption{Left panel: potential function~\eqref{eq:potential:garciabellido} considered in \Sec{sec:InflectionPoint:Plateau}, which features an inflection point (denoted with the vertical grey dashed line) at small field values and a plateau at large field values. The parameters are set to $\lambda=1$, $\xi=2.3$, $\alpha=6 \lambda \phi_\uc/(3 + \xi^2 \phi_\uc^4) - 4.3\times 10^{-5}$ and $m^2=\lambda \phi_\uc^2 (3 + \xi \phi_\uc^2)/(3 + \xi^2 \phi_\uc^4)$, where $\phi_\uc=0.64\Mp$ is the location of the inflection point. 
    Right panel: first and second slow-roll parameters as a function of the field value. Close to the inflection point, $\epsilon_2$ drops below $-3$ and reaches values close to $-6$, which signals the onset of a USR phase (see main text).
 \label{fig:inflectionpoint:potential}}
\end{figure}
Let us now investigate a model with a smoother inflection point, such as the one proposed in \Refs{Garcia-Bellido:2017mdw,Ezquiaga:2018gbw} where the potential function is given by
\bea 
\label{eq:potential:garciabellido}
V(\phi) = \frac{1}{12}\frac{6m^2\phi^2 - 4\alpha\phi^3+3\lambda\phi^4}{\left( 1+\xi\phi^2\right)^2} \, .
\eea 
It features an inflection point at small field values, and approaches a constant at large field values, such that the scales observed in the CMB emerge when the potential has a plateau shape, in agreement with CMB observations~\cite{Akrami:2018odb}. The potential function is displayed in the left panel of \Fig{fig:inflectionpoint:potential}. 

In the right panel, we display the first and second slow-roll parameters obtained from solving the Klein-Gordon equation~\eqref{eq:kleingordon} numerically. Let us recall that the first slow-roll parameter was introduced below \Eq{eq:KG:efolds} and is defined as $\epsilon_1=-\dot{H}/H^2$, while the second slow-roll parameter is defined as $\epsilon_2=\dd\ln\epsilon_1/\dd N$. At large field values, the system first undergoes a phase of attractor slow-roll inflation, where both $\epsilon_1$ and $\epsilon_2$ are small, which guarantees that the result does not depend on our choice of initial conditions (if taken sufficiently high in the potential). When approaching the inflection point, the slow-roll parameters become sizeable, which signals that slow roll breaks down (but inflation does not stop since $\epsilon_1$ remains below one). This triggers a USR phase, where $\epsilon_1$ rapidly decays and $\epsilon_2$ reaches values close to $-6$. Past the inflection point, $\epsilon_1$ increases again and grows larger than one, at which point inflation stops.

\begin{figure} 
    \centering
    \includegraphics[width=0.49\textwidth]{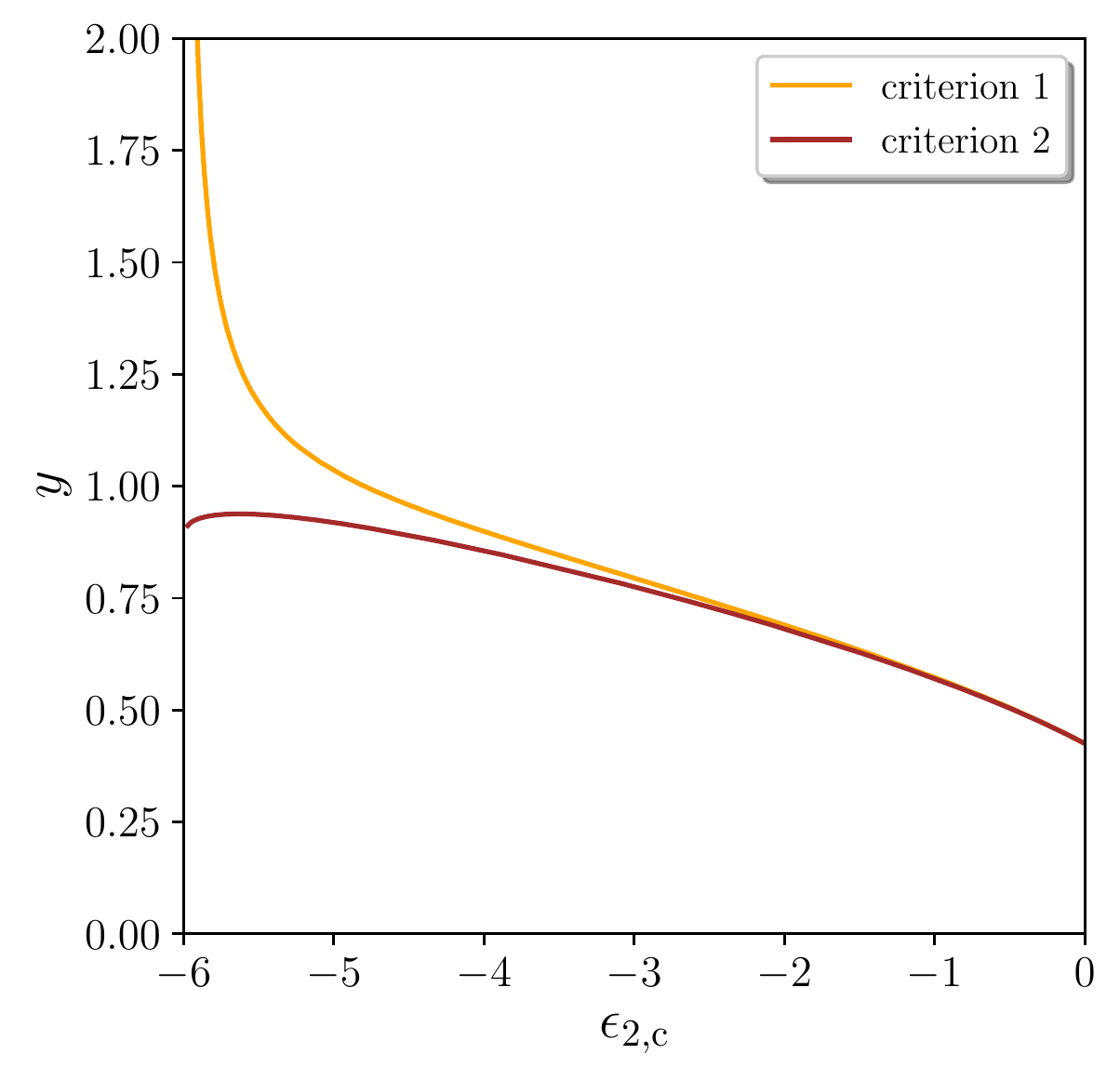}
    \includegraphics[width=0.49\textwidth]{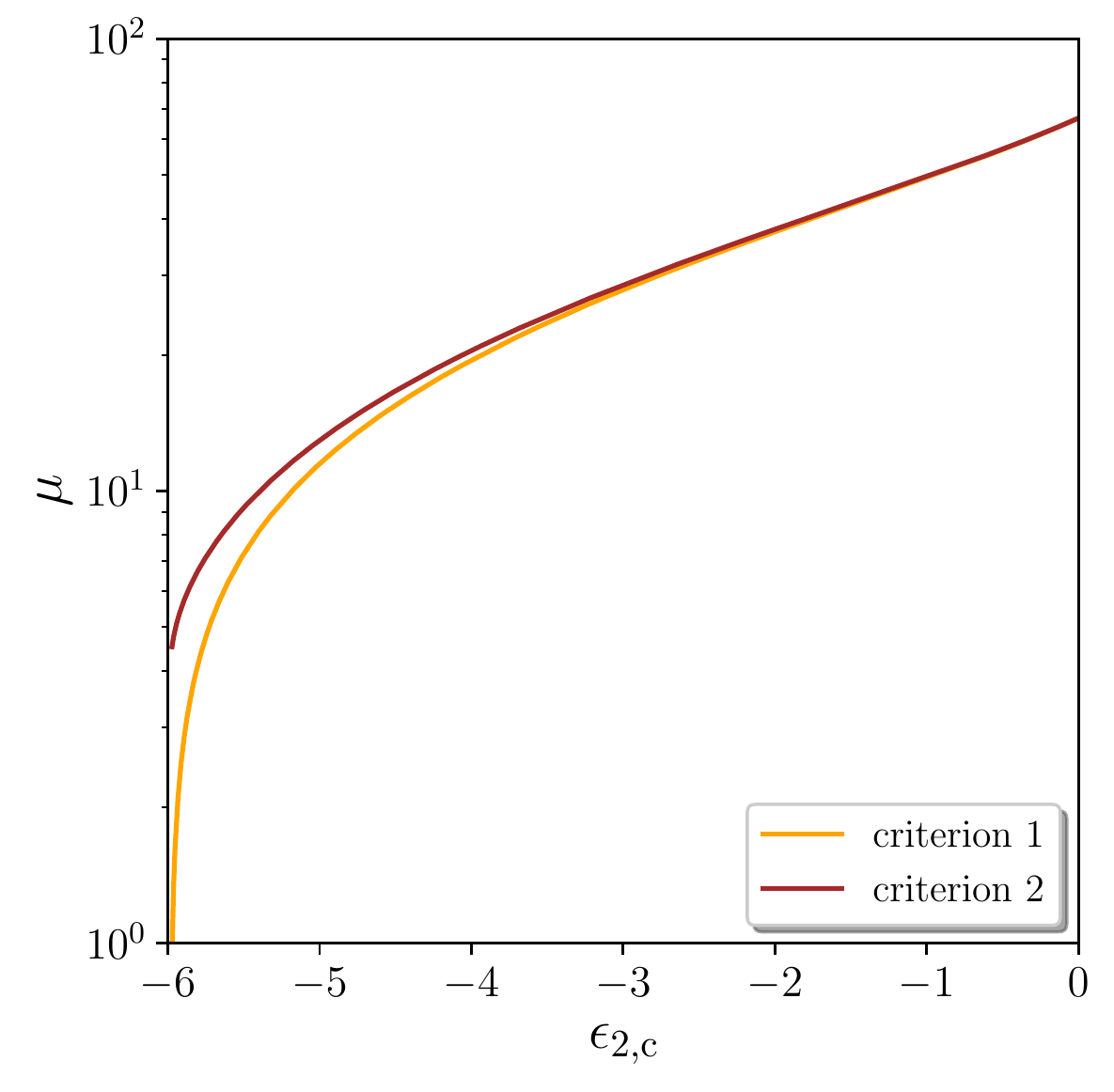}
    \caption{Effective $y$ and $\mu$ parameters for the inflection-point potential~\eqref{eq:potential:garciabellido}. The orange line, labeled ``criterion 1'', corresponds to defining USR as being when $\epsilon_2<\epsilon_{2,\uc}$. For the brown line, labeled ``criterion 2'', USR starts when $\epsilon_2<\epsilon_{2,\uc}$ and ends when $\epsilon_1$ reaches a minimum (hence $\epsilon_2$ becomes positive).
 \label{fig:inflectionpoint:ymu}}
\end{figure}
As explained in \Sec{sec:intro}, USR corresponds to when the acceleration term in the Klein-Gordon equation~\eqref{eq:KG:efolds}, $\ddot{\phi}$, dominates over the potential gradient term, $V'$. Using the definitions of the two first slow-roll parameters, one has~\cite{Pattison:2018bct}
\bea
\frac{V'}{\ddot{\phi}} = \frac{6}{2\epsilon_1-\epsilon_2}-1\, .
\eea 
As a consequence, USR, \ie $\vert \ddot{\phi} \vert > \vert V'\vert$, corresponds to $\epsilon_2<2\epsilon_1-3$, and maximal USR corresponds to when $\epsilon_2=2\epsilon_1-6$. This can be used to define the boundaries of the USR well, hence the effective parameters $y$ and $\mu$, which are displayed in \Fig{fig:inflectionpoint:ymu}. More precisely, one can define USR as being the regime where $\epsilon_2<\epsilon_{2,\uc}$, where $\epsilon_{2,\uc}$ is some critical value. This defines a certain region in the potential, the width and entry velocity of which can be computed, and \Eqs{eq:xy:variabletransforms} and~\eqref{eq:def:mu} give rise to the orange curves in \Fig{fig:inflectionpoint:ymu}. Strictly speaking, the definition of USR corresponds to taking $\epsilon_{2,\uc}=-3$ (if one neglects $\epsilon_1$). We however study the dependence of the result on the precise value of $\epsilon_{2,\uc}$, as a way to assess the robustness of the procedure which consists in investigating the inflection-point potential in terms of the toy model studied in \Secs{sec:StochasticInflation}-\ref{sec:volterra}. This procedure indeed implies a sharp transition towards USR, hence little dependence on the precise choice of the critical value $\epsilon_{2,\uc}$. For that same reason, in \Fig{fig:inflectionpoint:ymu}, we also display the result obtained by defining the USR phase in a slightly different way, namely USR starts when $\epsilon_2$ drops below $\epsilon_{2,\uc}$ and ends when $\epsilon_1$ reaches a minimum (hence when $\epsilon_2$ becomes positive). This is displayed with the brown line.

By definition, the two criteria match for $\epsilon_{2,\uc}=0$, but one can see that they give very similar results even for smaller values of $\epsilon_{2,\uc}$, and except if $\epsilon_{2,\uc}$ is very close to $-6$. This is a consequence of the abruptness with which USR ends. One can also see that the value of $y$ is always of order one, regardless of the value of $\epsilon_{2,\uc}$ and of the criterion being used (except if $\epsilon_{2,\uc}$ is very close to $-6$ and if one uses the first criterion). The reason is similar to the one advocated for the Starobinsky model in \Sec{sec:StarobinskyModel}: since the width of the USR well is defined as being where the classical trajectory relaxes to slow roll, by definition, the initial velocity of the inflaton is precisely the one such that the field reaches that point in finite time. 
The value of the $\mu$ parameter is always larger than one and depends more substantially on $\epsilon_{2,\uc}$, although it is never larger than $\mu\sim 50$. One may therefore expect the model to lie on the edge of the drift-dominated regime studied in \Sec{sec:classicallimit}, where the bulk of the PDF is quasi-Gaussian. 

However, given that the result produced by the toy model is very sensitive to the precise value of $\mu$, and that this value depends on the choice for $\epsilon_{2,\uc}$, one may question the applicability of the toy model in this case. Moreover, as explained in \Sec{sec:StarobinskyModel}, in a stochastic picture, USR does not end at a fixed field value, but $\epsilon_2(\bar{\phi},\bar{\pi})=\epsilon_{2,\uc}$ rather selects a certain hypersurface in phase space. The fact that we found USR ends abruptly in this model suggests that this may not be so problematic (namely that this hypersurface may have almost constant $\bar{\phi}$), but it is clear that our toy-model description can only provide qualitative results. A full numerical analysis seems therefore to be required to draw definite conclusions about this model. 

\section{Conclusions}
\label{sec:conclusions}

Single-field models of inflation able to enhance small-scale perturbations (e.g., leading to primordial black holes) usually feature deviations from slow roll towards the end of inflation, when stochastic effects may also play an important role. This is because, for large fluctuations to be produced, the potential needs to become very flat, such that the potential gradient may become smaller than the velocity inherited by the inflaton from the preceding slow-roll evolution, and possibly smaller than the quantum diffusion that vacuum fluctuations source as they cross out the Hubble radius.

In order to consistently describe the last stages of inflation in such models, one thus needs to formulate stochastic inflation, which allows one to describe the backreaction of quantum vacuum fluctuations on the large-scale dynamics, without assuming slow roll. More specifically, in the limit where the inherited velocity is much larger than the drift induced by the potential gradient, one enters the regime of ultra-slow roll. This is why in this work, we have studied the stochastic formalism of inflation in the ultra-slow-roll limit. 
As a specific example, we have considered a toy model where the potential is exactly flat over a finite region (which we dub the USR well), between two domains where the dynamics of the inflaton is assumed to be dominated by the classical slow-roll equations.

In this context, we have studied the distribution function of the first-exit time out of the well, which is identified with the primordial curvature perturbation via the $\delta N$ formalism. Contrary to the more usual, slow-roll case, the characteristic function of that distribution obeys a partial (instead of ordinary) differential equation, rendering the analysis technically more challenging. 

At the numerical level, we have presented results from simulations over a very large number of realisations of the Langevin equation. Although this method is direct, it is numerically expensive, and leaves large statistical errors on the tails of the distribution, where the statistics are sparse. This is why we have also shown how the problem can be addressed by solving Volterra integral equations, and designed an iterative scheme that provides a satisfactory level of numerical convergence as long as the inherited velocity of the field is not too small.

At the analytical level, we have studied the two limits where the inherited velocity is respectively much smaller and much larger than the required velocity to cross the well without quantum diffusion. In the small-velocity limit, at leading order we have recovered the slow-roll results obtained in \Refa{Pattison:2017mbe}. At higher orders, we have found that while the exponential decay rate of the first exit time distribution is not altered by the inherited velocity, its overall amplitude is diminished, such that the predicted abundance of primordial black holes is suppressed. In the large-velocity limit, the classical result is recovered at leading order, and we have presented a systematic expansion to compute corrections to the moments of the distribution function at higher order, that lead to results that are consistent with those previously found in \Refa{Firouzjahi:2018vet}. The tails of the distributions remain difficult to characterise analytically in this regime, as is the case for the classical limit in slow roll. However, the new approach we have proposed based on solving Volterra equations provides an efficient way to probe those tails numerically, which is otherwise very difficult to do with the numerical techniques commonly employed.  

We have then argued that this toy model can be used to approximate more realistic models featuring phases of ultra-slow roll. The idea is to identify the region in the potential where ultra-slow roll occurs, and approximate this region as being exactly flat (\ie neglecting the potential gradient). In practice, if one measures (i) the field width of that region, (ii) its potential height and (iii) the velocity of the inflaton when entering the flat region, one can use the results obtained with the toy model, that only depend on those three parameters, to get approximate predictions. This is similar to the approach employed (and tested) in \Refs{Pattison:2017mbe, Ezquiaga:2019ftu} for slow-roll potentials featuring a region dominated by quantum diffusion, where only the two first parameters are relevant.

Let us however highlight a fundamental difference with respect to the slow-roll case. In slow-roll potentials, it was found in \Refs{Vennin:2015hra, Pattison:2017mbe} that regions of the potential where stochastic effects dominate can be identified with the criterion $ v^2 \vert v'' \vert /(v')^2 \gg 1$ where $v=V/(24\pi^2\Mp^4)$. For a given potential function $v(\phi)$, this leads to well-defined regions in the potential, with well-defined field widths. In the ultra-slow roll case, the use of the toy model does not require that quantum diffusion dominates, but rather that the classical drift be much smaller than the inherited velocity. Since the velocity decays in time at a fixed rate, the transition from ultra-slow roll back to slow roll happens at a field value that a priori depends on time. Namely, at early time, the field velocity is still large and one needs a larger potential gradient to overtake it, while at larger times, once the field velocity has decayed, a smaller potential gradient is enough. In models where there is a sudden increase in the potential slope at a given field value, the transition always occurs close to that field value, but otherwise one may have to refine the present approach, which we plan to address in future work. 

\acknowledgments
We thank Jose Mar\'ia Ezquiaga, Hassan Firouzjahi, Juan Garc\'ia-Bellido, Amin Nassiri-Rad, Mahdiyar Noorbala, Ed Copeland, Cristiano Germani, and Diego Cruces for helpful discussions. 
HA and DW acknowledge support from the UK Science and Technology Facilities Council grants 
ST/S000550/1. 
Numerical computations were done on the Sciama High
Performance Compute (HPC) cluster which is supported
by the ICG, SEPNet and the University of Portsmouth.

\appendix 
\section{Drift-dominated limit and the method of characteristics}
\label{app:classicalLO:charfunction}
In this appendix, we solve the adjoint Fokker-Planck equation for the characteristic function, \Eq{eq:diff:chi}, in the regime where the dynamics of the system is dominated by the classical drift. At leading order, \Eq{eq:diff:chi} reduces to \Eq{eq:char:classicallimit}, which is a first order, linear, partial differential equation. It can thus be solved with the method of characteristics, which proceeds as follows. We first introduce the characteristic lines in phase space $\lbrace x(u), y(u) \rbrace$, where $u$ is an affine parameter along the line, and $x(u)$ and $y(u)$ are two functions that we want to choose such that $\chi_\mathcal{N}$ takes a simple form along the characteristic lines. By comparing
\bea
\label{eq:method_of_characteristic:chain_rule}
\frac{\dd}{\dd u}\chi_\mathcal{N}\left[t;x(u),y(u)\right] =\left( \frac{\dd x}{\dd u} \frac{\partial }{\partial x} + \frac{\dd y}{\dd u} \frac{\partial }{\partial y}\right)\chi_\mathcal{N}\left[t;x(u),y(u)\right]
\eea
with \Eq{eq:char:classicallimit}, one can see that $\dd\chi_\mathcal{N}/{\dd u}$ can be related to $\mathcal{L}_\mathrm{FP}^\dagger\cdot \chi_\mathcal{N}$ if one takes
\bea 
\frac{\dd x}{\dd u} = \frac{\dd  y}{\dd  u} = -3 \, ,
\eea 
which can be readily integrated as
\bea  \label{eq:characteristiccurves}
x &= x_0 -3u \, ,\quad
y &= y_0 - 3u \, ,
\eea 
where $x_0$ and $y_0$  are two integration constants that parametrise the various characteristic lines. Making use of  \Eq{eq:char:classicallimit}, \Eq{eq:method_of_characteristic:chain_rule} can thus be rewritten as
\bea 
\frac{\dd}{\dd u} \chi_\mathcal{N}\left[t;x(u),y(u)\right]= -\frac{it}{y_0 - 3u}\chi_\mathcal{N}\left[t;x(u),y(u)\right] \, ,
\eea 
which is solved by
\bea 
\label{eq:method_of_characteristic:chi:x0_y0}
\chi_\mathcal{N}\left[t;x(u),y(u)\right] = \chi_0(x_0,y_0)\left( y_0 - 3u  \right)^{\frac{it}{3}} \, , 
\eea 
where $\chi_0$ is an arbitrary function of the two parameters $x_0$ and $y_0$. Since phase space is two-dimensional, the characteristic lines can be labeled by a single parameter, here $x_0-y_0$, so we are free to set $y_0=0$. For a given point  $(x,y)$ in phase space, we thus have $x_0=x-y$ and $u=-y/3$, so in terms of the original variables $x$ and $y$, \Eq{eq:method_of_characteristic:chi:x0_y0} reads
\bea 
\chi_\mathcal{N}(t;x,y) = \chi_0\left( x-y \right)y^{\frac{it}{3}} \, .
\eea 
The function $\chi_0$ is set by imposing the boundary condition $\chi_{\mathcal{N}} (t; 0, y) = 1$, see \Eq{eq:char:initialconditions}, which yields $\chi_0(-y) = y^{\frac{-it}{3}}$, hence
\bea
\label{eq:char:classicalLO:App}
\left. \chi_\mathcal{N}\right\vert_\lo(t;x,y) = \left( 1 - \frac{x}{y}\right)^{-\frac{it}{3}} \, ,
\eea 
which is the solution given in \Eq{eq:char:classicalLO:solution}.

At next-to-leading order, the term proportional to $\partial^2/\partial x^2$ in \Eq{eq:diff:chi}, which we have neglected at leading order, can be evaluated with the leading-order solution~\eqref{eq:char:classicalLO:App}, leading to
\bea 
\label{eq:classical:NLO}
\left[ - 3y\left(\frac{\partial}{\partial x} + \frac{\partial}{\partial y}\right) + it \right]\left.\chi_{\mathcal{N}}\right\vert_\nlo(t;x,y)  = -  \frac{1}{y^2}\frac{it}{3\mu^2}\left(1+\frac{it}{3}\right)\left( 1- \frac{x}{y}\right)^{-2-\frac{it}{3}} \, ,
\eea 
see \Eq{eq:Drift_dominated:nlo:rec}. Using the same method of characteristics as before, along the same characteristic lines, the first-order partial differential equation can be solved again, and one obtains
\bea 
\label{eq:method_of_charactersitics:nlo:interm}
\left.\chi_{\mathcal{N}}\right\vert_\nlo(t;x,y) = \left[ \chi_1\left(y-x\right) + \frac{it}{9\mu^2}\left(1+\frac{it}{3}\right) \ln(y)\left( y-x\right)^{-2-\frac{it}{3}}\right]  y^{\frac{it}{3}} \, .
\eea 

In this expression, $\chi_1$ is a function of $x-y$ that needs to be determined by the boundary condition $\chi_{\mathcal{N}} (t; 0, y) = 1$, which leads to
\bea 
\label{eq:method_of_charactersitics:nlo:chi1}
\chi_1\left( y-x \right) = (y-x)^{-\frac{it}{3}}\left[ 1 - \frac{it}{9\mu^2 (y-x)^2}\left(1+\frac{it}{3}\right) \ln (y-x)   \right] .
\eea
Plugging \Eq{eq:method_of_charactersitics:nlo:chi1} into \Eq{eq:method_of_charactersitics:nlo:interm}, one obtains
\bea 
\left.\chi_{\mathcal{N}}\right\vert_\nlo(t;x,y) &= \left( 1 - \frac{x}{y}\right)^{-\frac{it}{3}}\left[ 1 - \frac{it}{9\mu^2}\left(1+\frac{it}{3}\right)\frac{\ln\left(1-\frac{x}{y}\right)}{(y-x)^2} \right] ,
\eea 
which is the formula given in \Eq{eq:characteristic:NLO}.
\section{Characteristic equation in separable form} \label{app:separable:charfunction}
In this appendix, we show how to solve the characteristic equation~\eqref{eq:diff:chi},
\bea
\label{eq:pde:chi:xy}
\left[\frac{1}{\mu^2} \frac{\partial^2}{\partial x^2}-3y\left(\frac{\partial}{\partial x} + \frac{\partial}{\partial y}\right)+i t \right]\chi_\mathcal{N}(t;x,y)=0\, ,
\eea 
through an expansion in the classical velocity $y$. This allows us to approach the solution in the diffusion-dominated regime discussed in \Sec{sec:stochasticlimit}.

The solution of the partial differential equation for the characteristic function, \Eq{eq:pde:chi:xy}, is not in general separable in terms of the variables $x$ and $y$. However, we can put the equation into a separable form by introducing variables corresponding to the characteristics of the differential equation, namely
\bea
\label{eq:rs:xy}
&r  =  x-y \,, & s =  y \,.
\eea
One recalls that, classically, $r$ is a constant. For $r<0$, the field has sufficient (negative) velocity to reach $\phi_\mathrm{end}$ classically, while for $r>0$ it would never reach $\phi_\mathrm{end}$ classically and this can only be achieved through quantum diffusion. The Langevin equation~\eqref{eq:langevin:xy} in terms of $r$ and $s$ is given by
\bea
 \label{eq:langevin:rs}
\frac{\dd r}{\dd N} &=  \frac{\sqrt{2}}{\mu} \xi(N)\, , \\
\frac{\dd s}{\dd N} &= -3s \, ,
\eea 
which shows that the dynamics of $r$ and $s$ decouple.
By rewriting \Eq{eq:pde:chi:xy} in terms of the new variables $r$ and $s$, we find 
\bea
\left( \frac{1}{\mu^2}\frac{\partial^2}{\partial r^2} - 3s\frac{\partial}{\partial s} + it \right)\chi_\mathcal{N}(t;r,s) = 0 \, .
\eea 
The solution for this partial differential equation can thus formally be written in a sum-separable form
\bea
\label{eq:separablechi}
\chi_\mathcal{N}(t;r,s) = \sum_n s^n R_n(t;r) \,,
\eea
where $R_n(t;r)$ is a solution of the oscillator equation
\bea
\frac{\partial^2}{\partial r^2} R_n(t;r) = - \omega_n^2(t) R_n(t;r) \,,
\eea
where $\omega_n^2(t)$ is the complex frequency
\bea
\label{eq:omegan}
\omega_n^2(t) = \left( it - 3n \right) \mu^2 
\,.
\eea

The price that we pay for simplifying the partial differential equation is that the boundary conditions \eqref{eq:char:initialconditions} at fixed $x$ now represent moving boundaries in our new variables. 
The boundary conditions are now given by 
\bea
\label{eq:rsbcs}
\left. \chi_\mathcal{N}(t; r, s) \right|_{r+s=0} = 1 \, , \quad 
\left. \frac{\partial \chi_\mathcal{N}}{\partial r}(t; r, s)  \right|_{r+s=1} = 0 \, ,
\eea 
and, in general, this leads to mode mixing in the apparently simple solution \eqref{eq:separablechi}.

It is interesting to notice that, at late time, $s$ (hence the classical velocity) goes to zero, so the diffusion-dominated limit can also be seen as a late-time limit. In this limit, only the first term can be kept in \Eq{eq:separablechi}, \ie, $\chi_\mathcal{N}(t;r,0)=R_0(t;r)$, and the boundary conditions \eqref{eq:rsbcs} yield a simple boundary condition in terms of $r$ for the $n=0$ mode with $\omega_0^2=it\mu^2$ and no mixing with other modes.
In fact, in the late-time limit, the solution of \Eq{eq:separablechi} is given exactly by \Eq{eq:chi:usr:latetime}, in agreement with both the slow-roll result and the late-time limit of the stochastic expansion detailed in \Sec{sec:stochlimit:expansion}. 

More generally, we write the mode function for each $n$ as
\bea
\label{eq:mode:n}
R_n(t;r) = A_n(t) \cos\left[ (1-r) \omega_n(t)\right] + B_n(t) \sin \left[ (1-r) \omega_n(t)\right] \, ,
\eea
and the boundary conditions \eqref{eq:rsbcs} can be written in terms of the mode sum \eqref{eq:separablechi} at a finite value of $s$ as
\bea
\label{eq:coefficientsAandB}
& \sum_n  s^n \left\{ 
A_n \cos \left[ (1+s)\omega_n\right] - B_n \sin \left[(1+s)\omega_n\right]
\right\} = 1
  \,,  \\
& \sum_n \omega_n s^n \left[ A_n \sin (\omega_n s) - B_n \cos(\omega_n s) \right] = 0 \,.
\eea 
By expressing $\sin (\omega_n s)$ and $\cos(\omega_n s)$ as power series in $s$, we see that these boundary conditions impose recurrence relations between the mode coefficients $A_n$ and $B_n$ for each power of $s$ in the overall sum \eqref{eq:separablechi}. 
Starting from the late-time limit, and calculating these coefficients up to second order, we find
\bea \label{eq:series:coefficients}
 A_0 &= \frac{1}{\cos\omega_0} \,, \hspace{1cm}  B_0 =0 \,, \ \\
 A_1 &= \frac{\omega_0}{\omega_1}  \frac{\omega_1\sin\omega_0 - \omega_0\sin\omega_1}{\cos\omega_0\cos\omega_1}
 \,,
 \hspace{1cm} B_1 =  \frac{\omega_0^2}{\omega_1} A_0
 \,, ~~~~~~~~  \\
A_2 &= 
 \frac{  \omega_1\sin\omega_1 A_1 - \omega_1\cos\omega_1 B_1-\sin\omega_2 B_2 + \frac{\omega_0^2}{2}\cos\omega_0 A_0}
 {\cos\omega_2} 
 \,,
 \hspace{1cm} B_2 = \frac{\omega_1^2}{\omega_2} A_1 \,. \ \ 
 \eea
 It is straightforward to compute the $A_n$ and $B_n$ coefficients up to any order that is desired, but the expressions become quickly cumbersome so we stop at second order here.

One can verify that the leading and next-to-leading order solutions derived in \Sec{sec:stochasticlimit} using a different method are properly recovered. At leading order, the late-time limit $\chi_{\mathcal{N}}(t;x,0)$ as $y\to0$ given in \Eq{eq:chi:usr:latetime} indeed corresponds to $\chi_{\mathcal{N}}(t;x,0)\to R_0(t;x)$ as $r\to x$ and $s\to 0$, corresponding to the lowest-order term in the sum \eqref{eq:separablechi}. Similarly one can recover the the first-order correction~\eqref{eq:solution:ftx} in the linear approximation for small $y$, \Eq{eq:linear:chi}, from the two lowest-order terms in the in the sum \eqref{eq:separablechi}
\bea
\chi_{\mathcal{N}}(t;r,s)\approx R_0(t;r) + s R_1(t;r)
\,,
\eea
substituting in for $r$ and $s$ in terms of $x$ and $y$ using \Eq{eq:rs:xy} and then keeping only the terms at zeroth and first orders in $y$.
One can see that the form of the relations \eqref{eq:coefficientsAandB} between the coefficients $A_n(t)$ and $B_n(t)$ in the mode functions \eqref{eq:mode:n} leads to a series of simple poles in the characteristic function \eqref{eq:separablechi}, whenever
\bea
\cos\left[\omega_m(t)\right]=0 \,,
\eea
and hence 
\bea
\omega_m(t) = \left( n+\frac12 \right) \pi \,,
\eea
where $n$ is an integer. 
The frequencies, $\omega_m(t)$, are given by \Eq{eq:omegan}. 
Hence we can write the characteristic function as
\bea 
\label{eq:pole:expansion:nm}
\chi_\mathcal{N}(t;r,s) = \sum_{n,m} \frac{a_{n}^{(m)}(r,s)}{\Lambda_{n}^{(m)}-it} + g(t;r,s) \, , 
\eea 
where $g(t;r,y)$ is some regular function, and $a_{n}^{(m)}(x,y)$ is the residual associated to the pole
$it=\Lambda_{n}^{(m)}$, where
\bea
\Lambda_{n}^{(m)} = 3m + \frac{\pi^2}{\mu^2}\left( n+\frac12 \right)^2 \,. 
\eea
\bibliographystyle{JHEP}
\bibliography{USRStochastic.bib}
\end{document}